# Modular Design of Hexagonal Phased Arrays through Diamond Tiles

Paolo Rocca, *Senior Member, IEEE*, Nicola Anselmi, *Member, IEEE*, Alessandro Polo, *Member, IEEE*, and Andrea Massa, *Fellow, IEEE*

***Abstract*—The modular design of planar phased array antennas with hexagonal apertures is addressed by means of innovative diamond-shaped tiling techniques. Both tiling configuration and sub-array coefficients are optimized to fit user-defined power-mask constraints on the radiation pattern. Towards this end, suitable surface-tiling mathematical theorems are customized to the problem at hand to guarantee optimal performance in case of low/medium-size arrays, while the computationally-hard tiling of large arrays is yielded thanks to an effective integer-coded *GA*-based exploration of the arising high-cardinality solution spaces. By considering ideal as well as real array models, a set of representative benchmark problems is dealt with to assess the effectiveness of the proposed architectures and tiling strategies. Moreover, comparisons with alternative tiling architectures are also performed to show to the interested readers the advantages and the potentialities of the diamond sub-arraying of hexagonal apertures.**

## I. Introduction

NOWADAYS, a key factor driving the evolution of phased array is the cost reduction [1] since services and products based on such a technology are expected to be widely diffused on the commercial market because of the constantly increasing demand of 5G applications (e.g., high-speed mobile communications, internet-of-things, and industry 4.0) and the boost of autonomous driving systems [2]. Indeed, phased arrays are considered an enabling technology in these fields thanks to their capability of guaranteeing the necessary quality-of-service and a suitable level of safety and reliability. On the other hand, even though the continuous development of electronics (e.g., fast analog-to-digital converters and massive systems-on-chip) and material science (e.g., artificial materials and meta-materials) as well as the introduction of innovative manufacturing processes (e.g., 3D printing and flexible electronics), phased arrays are still far from being commercial-off-the-shelf (*COTS*) devices. In recent years, innovative unconventional architectures have been studied and developed to provide better cost/performance trade-offs [3]. For instance, sparse [4]-[7] and thinned [8]-[11] arrays have been proposed to reduce the number of transmit-receive modules (*TRM*s) or radio-frequency (*RF*) chains that represent one of the main costs in phased arrays of modern radar systems [1]. These solutions have shown to be quite effective in yielding low sidelobes for interference and noise rejection and to afford arbitrary pattern shapes. However, sparse and thinned arrays cannot be implemented with a modular architecture and each design and manufacturing turns out to be *ad-hoc*. Furthermore, the final arrangement in sparse arrays is based on a non-uniform placement of the array elements within the antenna aperture and this implies an inefficient use of the space at disposal and a consequent directivity loss. Otherwise, the antenna aperture is fully exploited when dealing with sub-arrayed arrays in which the array elements, arranged on a regular lattice, are grouped and controlled at the sub-array level. The main issue of those unconventional architectures is the unavoidable presence of quantization and grating lobes [12][13] that become more and more impacting when increasing the scanning angle from the broadside direction and/or enlarging the operational bandwidth [14][15]. To cope with these drawbacks, several strategies have been presented for the optimization of the sub-array configuration and the sub-array amplitudes and/or phase excitation coefficients. Fast local-search techniques like the Contiguous Partition Method (*CPM*) [16][17][18] and the K-means [19] as well as global nature-inspired optimization algorithms [20][21][22][23] have been profitably proposed. All rely on the exploitation of unequal and arbitrary sub-array sizes and shapes to break the periodicity of the aperture illumination that causes the undesired quantization and grating lobes. More recently, tiled arrays have been also adopted [24][25][26][27][28]. They are clustered architectures where the sub-array modules belong to a finite set of simple tile shapes. The irregularity of the sub-array configuration, which allows the reduction of the undesired lobes, is here

Manuscript received March XX, 2019; revised October XX, 2019
This work has been partially supported by the Italian Ministry of Foreign Affairs and International Cooperation, Directorate General for Cultural and Economic Promotion and Innovation within the SNATCH Project (2017-2019) and by the Italian Ministry of Education, University, and Research within the Program "Smart cities and communities and Social Innovation" (CUP: E44G14000060008) for the Project "WATERTECH - Smart Community per lo Sviluppo e l'Applicazione di Tecnologie di Monitoraggio Innovative per le Reti di Distribuzione Idrica negli usi idropotabili ed agricoli" (Grant no. SCN_00489) and within the Program PRIN 2017 for the Project "CYBER-PHYSICAL ELECTROMAGNETIC VISION: Context-Aware Electromagnetic Sensing and Smart Reaction (EMvisioning)".

P. Rocca, N. Anselmi, A. Polo, and A. Massa are with the ELEDIA@UniTN (DISI - University of Trento), Via Sommarive 9, 38123 Trento - Italy (e-mail: {paolo.rocca, nicola.anselmi.1, alessandro.polo.1, andrea.massa}@unitn.it).
P. Rocca is also with the ELEDIA Research Center (ELEDIA@XIDIAN - Xidian University), 3P.O. Box 191, No.2 South Tabai Road, 710071 Xi'an, Shaanxi Province - China (e-mail: paolo.rocca@xidian.edu.cn)
A. Massa is also with the ELEDIA Research Center (ELEDIA@L2S - UMR 8506), 3 rue Joliot Curie, 91192 Gif-sur-Yvette - France (e-mail: andrea.massa@l2s.centralesupelec.fr)
A. Massa is also with the ELEDIA Research Center (ELEDIA@UESTC - UESTC), School of Electronic Engineering, Chengdu 611731 - China (e-mail: andrea.massa@uestc.edu.cn)
A. Massa is also with the ELEDIA Research Center (ELEDIA@TSINGHUA - Tsinghua University), 30 Shuangqing Rd, 100084 Haidian, Beijing - China (e-mail: andrea.massa@tsinghua.edu.cn)

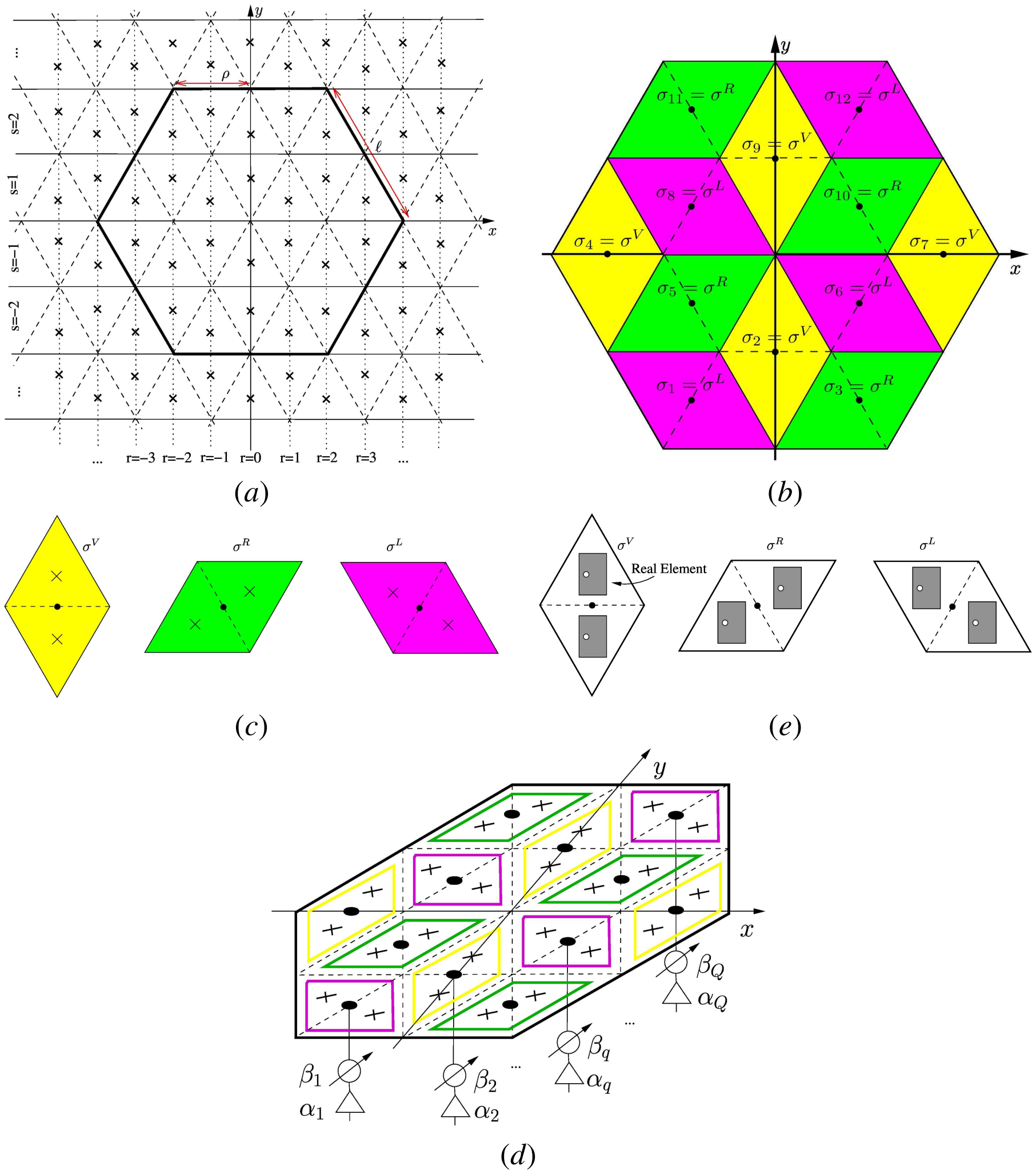

Figure 1. *Illustrative Geometry* ($N_\ell = 2$, $N = 24$, $Q = 12$) - Sketch of ($a$) a hexagonal aperture on a honeycomb lattice structure, ($b$) a tiling configuration with ($c$) the three diamond-shaped tiles (vertical, $\sigma^V$, horizontal-left, $\sigma^L$, and horizontal-right, $\sigma^R$, orientations), and ($d$) the corresponding sub-arrayed beam-forming network. A realistic model of the three tiles with two patch antenna elements ($e$).

obtained by properly choosing suitable tile shapes and by optimizing their disposition and orientation within the aperture. For instance, polyomino-shaped sub-arrays have been used in [24] and a customized Genetic Algorithm (*GA*) has been also proposed to maximize the aperture coverage as well as the operational bandwidth for a given aperture size [26]. Thanks to the exploitation of pre-defined sub-array shapes, tiled apertures are simple and modular structures, thus potential candidate solutions for synthesizing low-cost and large-scale/mass production phased arrays. Unfortunately, the literature on optimal-design methodologies for array tiling, which enable the full coverage of the array aperture with given tile shapes, is still quite scarce. This is due to the mathematical complexity of the clustered array design [29] and the limited set of combinations between tiles- and aperture-shapes [30][31][32][33] for which an optimal solution has been found. In [34], a mathematical theory for clustering rectangular apertures with domino-shaped tiles has been customized to the design and optimal tiling of phased array antennas. Pertinent theorems from [35][36], which assure a complete tessellation of the antenna aperture and the generation of all possible tiling configurations, have been properly exploited. This work extends such a theoretical framework to the modular design of phased arrays with hexagonal aperture through an irregular and optimal placement of diamond-shaped tiles. Starting from the mathematical theory on the tessellation of hexagonal surfaces, two strategies, one enumerative and one computational, are derived for the optimization of the tiling configuration and the sub-array coefficients to fit user-defined power-mask constraints on the power pattern radiated by regular hexagonal apertures. The interest in phased array with hexagonal aperture is high since they are more suitable, with respect to rectangular arrays, for scanning the mainlobe around a pointing direction due to their six-fold symmetry and since they better fit circular shapes, as well. Moreover, the hexagon is one of the three regular polygons, along with the square and the equilateral triangle, that allows the perfect tessellation of the Euclidean plane and it can be easily combined to compose complex planar/conformal structures (e.g., geodesic domes) [38][39]. As for the main novelties of this research work over the existing literature on the field and to the best of the authors' knowledge, they comprise: (*i*) the engineering exploitation of mathematical theorems providing the conditions for the full coverage of hexagonal apertures with diamond-shaped tiles [40][41][42] and the knowledge of the total number of tiling configurations [43][44]; (*ii*) the introduction of an innovative enumerative strategy, which is based on analytic rules [45], previously applied to the case of domino-tiled arrays [34] but here derived for diamond tiles, to determine the optimal tiling configuration in case of low- and medium-size arrays; (*iii*) the exploitation of an innovative customization of an Integer-coded *GA* (*IGA*), while a standard binary *GA* (*BGA*) has been used in [34], for an effective exploration of high-cardinality solution spaces to enable the design of large arrays affording mask-constrained power patterns.

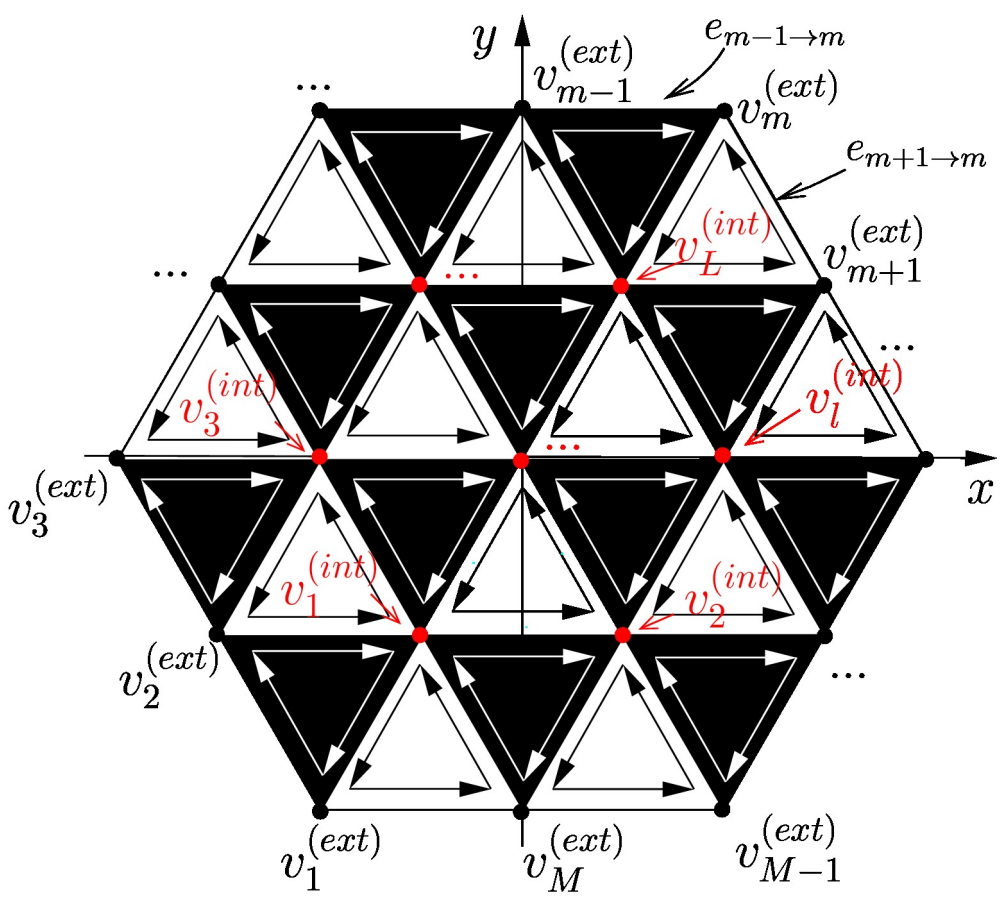

Figure 2. *Illustrative Geometry* ($N_\ell = 2$, $N = 24$) - Black-and-white representation of the triangular unit cells composing the hexagonal array aperture $A$ with external vertices, $\{v_m^{(ext)};\, m = 1, ..., M\}$, internal vertices, $\{v_l^{(int)};\, l = 1, ..., L\}$, and pixel-edges, $\{e_{m\to(m\pm1)};\, m = 0, ..., M-1\}$.

The rest of the paper is organized as follows. The mathematical formulation of the tiling problem of hexagonal-aperture phased arrays with diamond-shaped tiles is reported in Sect. 2. Section 3 is devoted to the description of the synthesis strategies for low/medium and large arrays. The validation and the comparative assessment of the proposed tiling methods are carried out in Sect. 4 by means of a set of representative numerical examples concerned with realistic array models, as well. Eventually, some conclusions and final remarks are drawn (Sect. 5).

## II. Mathematical Formulation

Let us consider a planar phased array defined over a regular hexagonal aperture $A$ ($l_i = \ell$, $i = 1, ..., 6$, $l_i$ being the length of its $i$-th side) where $N$ elementary radiators are displaced on the $xy$-plane according to the honeycomb lattice in Fig. 1($a$), $\rho$ being the side of its unit cell equal to an equilateral triangle. With reference the $1^{st}$ Quadrant ($x \geq 0$, $y \geq 0$), the values of the coordinates of the lattice points (i.e., the candidate positions of the barycenters of the array elements) are

$$x_r = r \times \frac{\rho}{2} \tag{1}$$

and

$$y_s = \begin{cases} \left(\frac{3s-2}{\sqrt{3}}\right) \times \frac{\rho}{2} & if\ \ (r\ and\ s)\ are\ both\ even\ or\ odd \\ \left(\frac{3s-1}{\sqrt{3}}\right) \times \frac{\rho}{2} & otherwise \end{cases} \tag{2}$$

where $(r, s)$ is a couple of integer indexes that univocally identifies a point of the lattice, $r$ ($r \geq 0$) and $s$ ($s \geq 1$) being the column and the row-strip index, respectively [Fig. 1($a$)]. For the sake of symmetry, the positions of the other barycenters belonging to the other quadrants can be easily inferred by means of mirroring operations with respect to the $x$ and $y$ axes: $x_{-r} = -x_r$ and $y_s = y_s$ ($2^{nd}$ Quadrant), $x_{-r} = -x_r$ and $y_{-s} = -y_s$ ($3^{rd}$ Quadrant), and $x_r = x_r$ and $y_{-s} = -y_s$ ($4^{th}$ Quadrant). The array antenna is composed by $Q$ clusters, $\{\sigma_q;\, q = 1, ..., Q\}$, each one grouping two adjacent unit cells of $A$ (i.e., $Q = \frac{N}{2}$) that share one edge [Fig. 1($b$)] so that the resulting tiles have a *diamond* shape and three possible orientations: vertical, $\sigma^V$, horizontal-left, $\sigma^L$, and horizontal-right, $\sigma^R$ [Fig. 1($c$)]. While the fabrication of a single tile and its rotation to obtain $\sigma^V$, $\sigma^L$, and $\sigma^R$ could be a mathematically viable solution, it is not admissible from an electromagnetic viewpoint, thus three different primitives/building-blocks, ($\sigma^V$, $\sigma^L$, $\sigma^R$), are necessary. For instance, let the elementary radiator be a patch antenna with linear or horizontal polarization. To avoid any polarization issues on the radiated electromagnetic (*EM*) field, the three tiles sketched in Fig. 1($e$) must be implemented.

For a given tiling configuration [e.g., Fig. 1($b$)], the signal either transmitted or received to/from each $q$-th ($q = 1, ..., Q$) tile is controlled by a *TRM* characterized by an amplitude coefficient $\alpha_q$ and a phase delay $\beta_q$ [Fig. 1($d$)], while the *EM* field generated in far-field is given by

$$\mathbf{E}(u, v) = \sum_{s=-N_\ell,\, s\neq 0}^{N_\ell} \sum_{r=-(2\times N_\ell - |s|)}^{2\times N_\ell - |s|} \mathbf{p}_{rs}(u, v) \times \sum_{q=1}^{Q} \delta_{c_{rs}q}\alpha_q e^{j[k(x_r u + y_s v) + \beta_q]} \tag{3}$$

where $\mathbf{p}_{rs}(u, v)$ is the embedded element pattern [12][13] of the $(r, s)$-th element of the honeycomb lattice of the array, $N_\ell$ ($N_\ell \triangleq \frac{\ell}{\rho}$) is the number of unit cells that exist on each side of $A$, and $N^{(s)}$ ($N^{(s)} \triangleq 4 \times N_\ell - 2 \times |s| + 1$) is the total number of elements within the $s$-th row strip [i.e., $\sum_{s=-N_\ell,\, s\neq 0}^{N_\ell} N^{(s)} = N$]. Moreover, $k = \frac{2\pi}{\lambda}$ is the wavenumber, $\lambda$ being the wavelength, $u = \sin\theta\cos\phi$ and $v = \sin\theta\sin\phi$ are the direction cosines ($\theta \in [0 : 90]$ [deg] and $\phi \in [0 : 360]$ [deg]). Furthermore, $\mathbf{c}$ is the membership vector ($\mathbf{c} = \{c_{rs};\, r = -\left(\frac{N^{(s)}-1}{2}\right), ..., \left(\frac{N^{(s)}-1}{2}\right);\, s = -N_\ell, ..., N_\ell;\, s \neq 0\}$) whose $(r, s)$-th entry is an integer value ($c_{rs} \in [1 : Q]$) that identifies the membership of the $(r, s)$-th element of the array lattice to the $q$-th diamond tile (i.e., $c_{rs} = q$ if the $(r, s)$-th element belongs to the $q$-th tile), while $\delta_{c_{rs}q}$ is the Kronecker delta ($\delta_{c_{rs}q} = 1$ if $c_{rs} = q$ and $\delta_{c_{rs}q} = 0$ otherwise [$c_{rs} \neq q$]).

The degrees-of-freedom (*DoF*s) of the tiled array architecture at hand, namely the sub-array configuration, $\mathbf{c}$, and the amplitude, $\boldsymbol{\alpha}$ ($\boldsymbol{\alpha} = \{\alpha_q : q = 1, ..., Q\}$), and the phase, $\boldsymbol{\beta}$ ($\boldsymbol{\beta} = \{\beta_q : q = 1, ..., Q\}$), excitation vectors, are defined when solving the following "*mask-closeness*" synthesis problem:

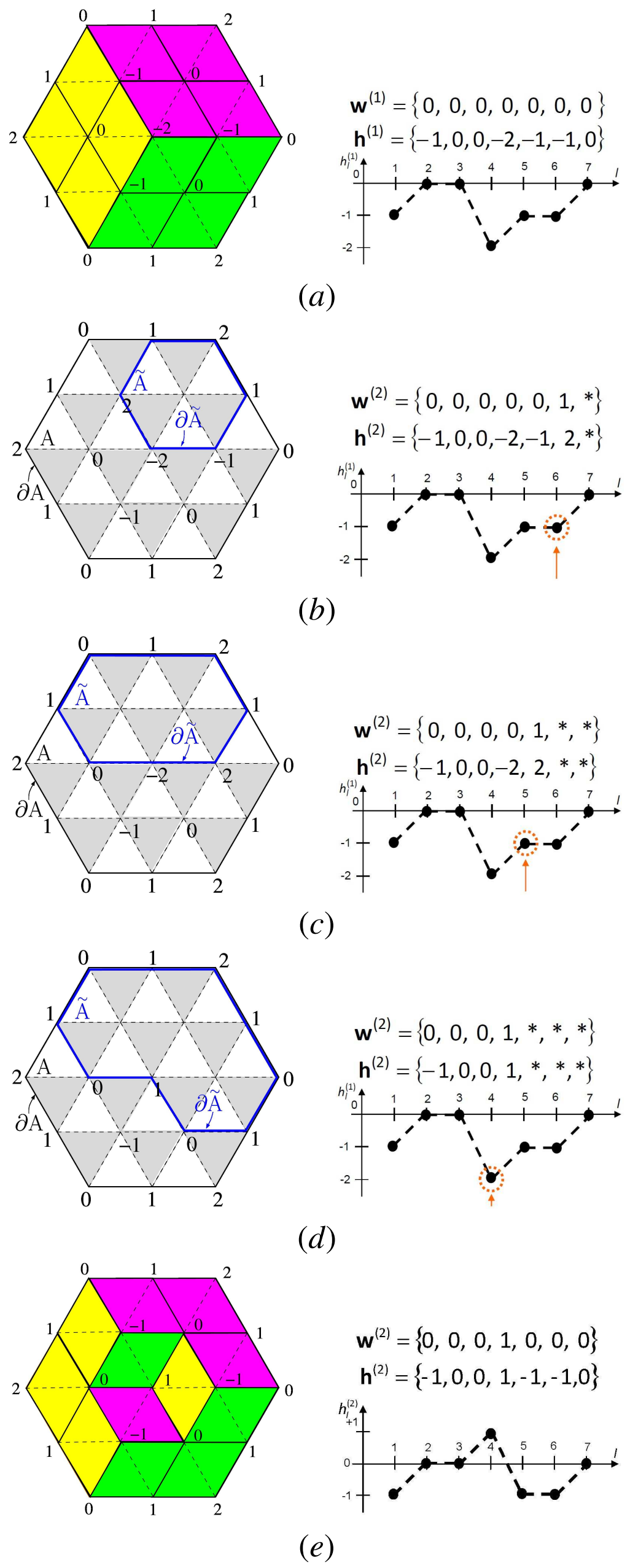

Figure 3. *EDM Synthesis* ($N_\ell = 2$, $N = 24$) - Generation of the second ($t = 2$) tiling word from the first/minimal ($t = 1$) one. *HF* values of the internal vertices, $\{v_l^{(int)};\, l = 1, ..., L\}$, and tiling words for $(a)$ the minimal tiling $\mathbf{c}^{(1)}$ (i.e., $\mathbf{w}^{(1)}$) and $(e)$ the second tiling $\mathbf{c}^{(2)}$ (i.e., $\mathbf{w}^{(2)}$) along with $(b)(c)(d)$ three intermediate solutions not satisfying $(b)(c)$ and satisfying $(d)$ the tileability condition (9).

*Diamond-Tiling Hexagonal Array Synthesis* - Given a fully-populated hexagonal array [Fig. 1$(a)$] fitting a user-defined power mask $U(u, v)$, find its optimal sub-arraying configuration $\mathbf{c}^{opt}$ (i.e., the optimal arrangement of vertical, $\sigma^V$, horizontal-left, $\sigma^L$, and horizontal-right, $\sigma^R$, tiles fully covering the aperture $A$) and the corresponding values of the sub-array amplitude, $\boldsymbol{\alpha}^{opt}$, and phase, $\boldsymbol{\beta}^{opt}$, coefficients so that the cost function $\chi$

$$\chi(\mathbf{c}, \boldsymbol{\alpha}, \boldsymbol{\beta}) \triangleq \int_\Omega \left[|\mathbf{E}(u, v)|^2 - U(u, v)\right] \times \mathcal{H}\left\{|\mathbf{E}(u, v)|^2 - U(u, v)\right\} dudv \tag{4}$$

is minimized, that is, the radiated pattern minimally violates the mask constraint, $\mathcal{H}\{\cdot\}$ and $\Omega$ being the Heaviside function and the visible range ($\Omega = \{(u, v) : u^2 + v^2 \le 1\}$), respectively.

## III. Hexagonal Array Diamond-Tiling Methodologies

In order to address the "*Diamond-Tiling Hexagonal Array Synthesis*" problem at hand (Sect. 2), two innovative design methodologies are presented in the following. Unlike [34], where domino tiles and rectangular apertures with the array elements located on a rectangular lattice have been taken into account, the proposed approaches, namely the Enumerative Design Method (*EDM*) and the Computational Design Method (*CDM*), are specific for the tiling of hexagonal arrays with diamond-shaped sub-arrays. Both methods benefit from the tileability theorem (*Appendix I*) [40][41][42], which states the conditions for the full-coverage with diamond tiles of hexagonal array apertures with radiating elements disposed on a honeycomb lattice. They are also based on the optimal tiling algorithm [45] that assures the exhaustive generation of all possible $T$ tiling configurations, $\{\mathbf{c}^{(t)},\ t = 1, ..., T\}$, (*Appendix II*) fitting the full coverage condition. Therefore, when one has "enough" time and/or computational resources to generate the whole set of $T$ admissible tiling configurations and to check the optimality of each $t$-th ($t = 1, ..., T$) solution by computing (4), the *EDM* allows one to faithfully retrieve the optimal problem solution [i.e., the global minimum of (4)], which corresponds to the best tiled-array performance. Otherwise (i.e., if testing all $T$ solutions becomes computationally intractable), the *CDM* enables an effective/computationally-efficient sampling of the tiling-solution space by means of a customized *IGA*. It is worth pointing out that the integer-coding of the *DoF*s is aimed at improving the efficiency and the convergence rate of the *GA*, with respect to the *BGA* used in [34], but also to extend the applicability of the analytic schemata-driven tiling optimization [34] to larger arrays. Since the choice of which methodology to adopt is of importance for an antenna designer and the use of one or the other methodology should not be left to chance, the analytic relationship derived from [43][44] for the cardinality $T$ of the solution space (*Appendix II*) can be *a-priori* exploited to estimate the *CPU*-time required by the *EDM* [i.e., $\tau = T \times \Delta t$, $\Delta t$ being the amount of time for generating a new trial solution and to compute (4)] and its admissibility with the array size at hand.

### A. Enumerative Design Method

Given a fully diamond-tileable aperture $A$ (*Appendix I*), the *EDM* generates and evaluates all $T$ tiling configurations, $\{\mathbf{c}^{(t)};\ t = 1, ..., T\}$, each one characterized by a different spatial arrangement of the diamond tiles $\sigma^V$, $\sigma^L$, and $\sigma^R$ [Fig. 1$(c)$]. Like the optimal tiling algorithm in [45], the *EDM* exploits the *Height Function* (*HF*) $h(\cdot)$ [36], which is defined on the vertices of the check-board pattern in Fig. 2 and its values are determined (*Appendix III*) by considering the corresponding edges, to univocally encode the $N$-dimensional tiling vector $\mathbf{c}^{(t)}$ into a smaller $L$-dimensional ($L \triangleq 3 \times N_\ell^2 - 3 \times N_\ell + 1$, $L \ll N$) string of integer values, called *tiling word* $\mathbf{w}^{(t)}$ ($t = 1, ..., T$). The tiling words are iteratively yielded starting from the first one, called *minimal tiling word*, having all entries equal to zero (i.e., $\mathbf{w}^{(t)}\big|_{t=1} = \mathbf{w}^{(1)} = \{w_l^{(1)} = 0;\, l = 1, ..., L\}$). More in detail, the generation of the second ($t = 2$) tiling word, $\mathbf{w}^{(2)}$, starts with the selection of the internal vertex with the largest index $\xi \in [1 : L]$ that satisfies one of the following conditions [Fig. 3($b$) - right plot, $\mathbf{h}^{(2)}$]

$$\begin{array}{ll} h_{\xi-1}^{(t-1)} \geq h_\xi^{(t-1)} \leq h_{\xi+1}^{(t-1)} & \xi \in [2 : L-1] \\ h_\xi^{(t-1)} \leq h_{\xi+1}^{(t-1)} & \xi = 1 \\ h_\xi^{(t-1)} \leq h_{\xi-1}^{(t-1)} & \xi = L. \end{array} \tag{5}$$

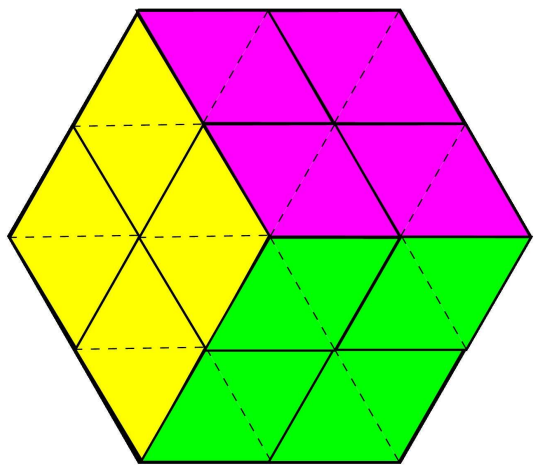

Figure 4. *EDM Synthesis* ($N_\ell = 2$, $N = 24$) - Sketch of the arrangement of vertical, $\sigma^V$, horizontal-left, $\sigma^L$, and horizontal-right, $\sigma^R$, diamond-tiles for the *minimal tiling,* $\mathbf{c}^{(1)}$.

Table I
*Solution-Space Cardinality* - NUMBER OF COMPLETE TILING CONFIGURATIONS, $T$, FOR A SET OF REPRESENTATIVE $N$-ELEMENTS ARRAYS WHEN USING DOMINO OR DIAMOND CLUSTERS.

| | *Hexagonal Aperture* | | *Rectangular Aperture* | |
|---|---|---|---|---|
| $N$ | $N_\ell$ | $T^{(hex)}$ | $N_x \times N_y$ | $T^{(dom)}$ |
| 24 | 2 | 20 | $6 \times 4$ | 281 |
| 54 | 3 | 980 | $6 \times 9$ | $8.1799 \times 10^5$ |
| 96 | 4 | $2.3285 \times 10^5$ | $8 \times 12$ | $8.2741 \times 10^{10}$ |
| 150 | 5 | $2.6723 \times 10^8$ | $10 \times 15$ | $2.6450 \times 10^{17}$ |
| 216 | 6 | $1.4786 \times 10^{12}$ | $12 \times 18$ | $2.9186 \times 10^{25}$ |
| 294 | 7 | $3.9406 \times 10^{16}$ | $14 \times 21$ | $1.0216 \times 10^{35}$ |

Then, the first $\xi - 1$ letters of $\mathbf{w}^{(t-1)}$ are copied into $\mathbf{w}^{(t)}$

$$w_j^{(t)} = w_j^{(t-1)} \quad j = 1, ..., \xi - 1, \tag{6}$$

while the $\xi$-th one is increased of one unity [Fig. 3($b$) - right plot, $\mathbf{w}^{(2)}$]

$$w_\xi^{(t)} = w_\xi^{(t-1)} + 1. \tag{7}$$

The *HF* values for the first $\xi$ internal vertices are computed as follows

$$h_j^{(t)} = 3 w_j^{(t)} + h_j^{(1)} \quad j = 1, ..., \xi \tag{8}$$

where $\{h_l^{(1)};\, l = 1, ..., L\}$ are the *HF* values of the *minimal tiling* configuration $\mathbf{c}^{(1)}$ (Fig. 4) computed as detailed in *Appendix IV*.
If the condition

$$\left| h_j^{(t)} - h_i^{(t)} \right| = \{1, 2\} \tag{9}$$

is not verified for every couple of neighboring vertices $v_i^{(t)}$ and $v_j^{(t)}$ on $\partial\tilde{A}$, $\partial\tilde{A}$ being the boundary of $\tilde{A}$, $\tilde{A}$ being the portion of the aperture $A$ in which the *HF* values of internal vertices have not been yet defined [Fig. 3($b$) - left plot], the second largest index $\xi \in [1 : L]$ satisfying (5) is taken into account [Fig. 3($c$) - right plot, $\mathbf{h}^{(2)}$] and the steps (6), (7), and (10) are carried out. Such a procedure is iterated until a value $\xi \in [1 : L]$ is found [Fig. 3($d$) - right plot, $\mathbf{h}^{(2)}$] so that (9) holds true [Fig. 3($d$) - left plot]. The diamond tiles are then placed in the region $\left(A - \tilde{A}\right)$ to match the tiling conditions (18). The new complete tiling configuration $\mathbf{c}^{(t)}$ and the corresponding *HF* values [Fig. 3($e$)] are determined by means of the *Thurston*'s algorithm [36] (*Appendix V*). Finally, the missing letters ($j = \xi + 1, ..., L$) of the new tiling word, $\mathbf{w}^{(t)}$, are computed as follows

$$w_j^{(t)} = \frac{h_j^{(t)} - h_j^{(1)}}{3}. \tag{10}$$

The previous procedure for defining $\mathbf{w}^{(2)}$ starting from $\mathbf{w}^{(1)}$ is then successively applied to generate $\mathbf{w}^{(t)}$ from $\mathbf{w}^{(t-1)}$ until $t = T$.
Once all tiling words, $\{\mathbf{w}^{(t)};\, t = 1, ..., T\}$ and the corresponding sub-array configurations, $\{\mathbf{c}^{(t)};\, t = 1, ..., T\}$, have been generated, the optimal tiling, $\mathbf{c}^{opt}$, is selected among the whole $T$ set as the diamond-tiled arrangement whose cost function value $\chi^{(t)}$ [$\chi^{(t)} \triangleq \chi\left(\mathbf{c}^{(t)}, \boldsymbol{\alpha}^{(t)}, \boldsymbol{\beta}^{(t)}\right)$] is minimum

$$\mathbf{c}^{opt} = \arg\left(\min_{t=1,...,T}\left\{\chi^{(t)}\right\}\right) \tag{11}$$

by setting the $t$-th sub-array vectors, $\boldsymbol{\alpha}^{(t)}$ and $\boldsymbol{\beta}^{(t)}$, to

$$\left(\begin{array}{c} \alpha_q^{(t)} \\ \beta_q^{(t)} \end{array}\right) = \frac{1}{2}\sum_{s=-N_\ell,\, s\neq 0}^{N_\ell} \sum_{r=-\left(\frac{N^{(s)}-1}{2}\right)}^{\left(\frac{N^{(s)}-1}{2}\right)} \delta_{c_{rs}q} \times \left(\begin{array}{c} \alpha_{rs}^{ref} \\ \beta_{rs}^{ref} \end{array}\right) \tag{12}$$

$\left(\alpha_{rs}^{ref}, \beta_{rs}^{ref}\right)$ being the $(r, s)$-th [$r = -\left(\frac{N^{(s)}-1}{2}\right), ..., \left(\frac{N^{(s)}-1}{2}\right)$; $s = -N_\ell, ..., N_\ell$; $s \neq 0$] reference amplitude and phase coefficients of a fully-populated array antenna (i.e., an array having an independent *TRM* with a dedicated amplifier and a phase shifter for each array element).

## B. Computational Design Method

When the number of array elements $N$ significantly grows, the synthesis with the *EDM* is unaffordable since the cardinality of the solution space $T$ turns out to be very high (Tab. I and Fig. 5) despite the use of $L$ integer values (i.e., the dimension of a word, $\mathbf{w}$, namely the number of word letters) to univocally encode a trial tiling solution instead of $N$ (i.e., the dimension of $\mathbf{c}^{(t)}$) ($\frac{L}{N} < \frac{1}{2}$) thanks to the bijective relationship between the *HF* values of the internal vertices and a trial tiling configuration (Fig. 3). The *CDM* exploits a novel *GA*-based optimization to enable the synthesis of large arrays through an effective sampling of the solution space and the convergence towards - or very close - to the global optimum. The main novelty with respect to the *GA* in [34] is the integer coding of the tiling words to considerably reduce the chromosome length and, in turn, to map the original solution space of dimension $N$ into a $L$-sized smaller one still of cardinality $T$ [$T = (N_\ell)^L$] to speed up the convergence and to enable the tiling of wider arrangements, as well. For example, let us suppose the maximum value of a letter in a word be $w_{max} = 4$ ($w_{max} \triangleq \max_{t=1,...,T;\, l=1,...,L}\left\{w_l^{(t)}\right\}$) and the word length is $L = 37$, when applying the *BGA*, the chromosome [34] has a length equal to 111 bits (Tab. II) since three bits are needed to encode the set of admissible values of a letter (i.e., $w_l^{(t)} = \{0, 1, 2, 3, 4\}$). Differently, the chromosome length of the *IGA* is exactly equal to $L$ (Tab. II). Furthermore, the *IGA* turns out to be more advantageous than the *BGA* for the following reasons: (*i*) the operation of coding the tiling words into binary chromosomes and decoding the chromosomes into tiling words is avoided, thus saving *CPU*-time; (*ii*) the direct use of integer variables (i.e., the genes of the chromosome) allows one to *a-priori* and intrinsically handle constraints on the admissible values of the word letters, which is not doable in the *BGA*. Indeed, it is known that the minimum value that a letter may assume is zero and it is that of the *minimal tiling* ($\left. w_l^{(t)}\right|_{t=1} = 0$; $l = 1, ..., L$), while the maximum one is equal to that of the *maximal tiling* ($\left. w_l^{(t)}\right|_{t=T}$; $l = 1, ..., L$). This latter corresponds to the "depth" of the internal vertex, namely, the minimum number of edges that, starting from an external vertex on $\partial A$, one has to cross for reaching the internal vertex. For illustrative purposes, two representative examples of the maximal tiling for $N_\ell = 2$ and $N_\ell = 5$ are shown in Fig. 6.

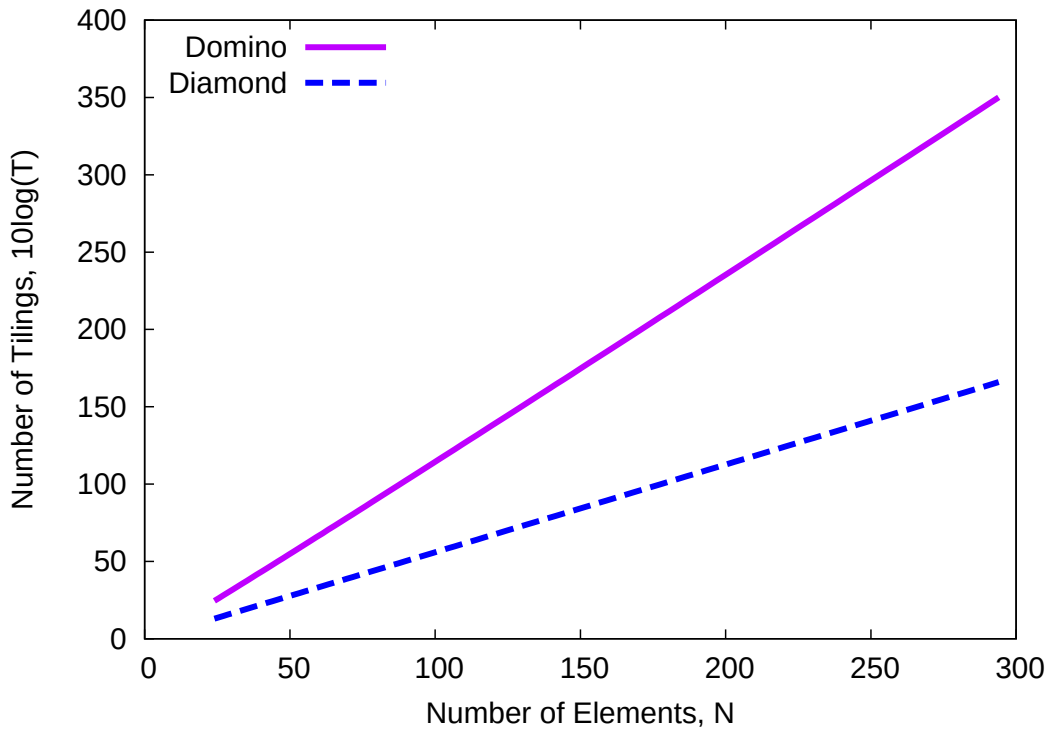

Figure 5. *Solution-Space Cardinality* - Number of complete tiling configurations, $T$, of an array composed by $N$ elements clustered with diamond or domino tiles.

Table II
*Illustrative Example* ($L = 37$) - CHROMOSOME SEQUENCE USING AN INTEGER-CODING (*IGA*) AND A BINARY-CODING (*BGA*).

| | *Chromosome* |
|---|---|
| *IGA* | 1111122211233211234321123321122211111 |
| *BGA* | 001001001001001010010010001001010011011010001001010011 1<br>0001101000100101001101101000100101001001000100100100 1001 |

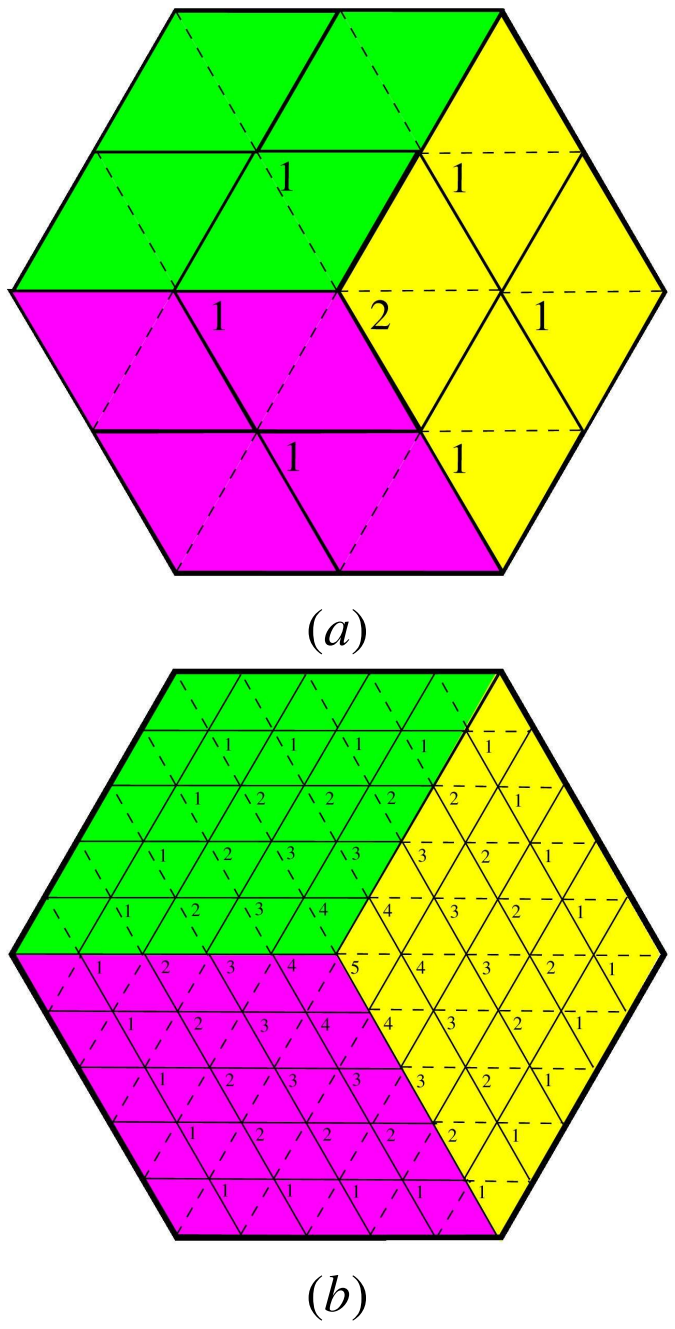

Figure 6. *CDM Synthesis* - *Maximal tiling*, $\mathbf{c}^{(T)}$, for a hexagonal array with ($a$) $N_\ell = 2$, $N = 24$ and ($b$) $N_\ell = 5$, $N = 150$.

As for the *IGA*, it follows the algorithmic implementation of the optimization method described in [34]. Thanks to the possibility of *a-priori* and analytically defining the first/minimal tiling word $\mathbf{w}^{(1)}$ and the last/maximal tiling word $\mathbf{w}^{(T)}$ along with the constraint that

$$w_l^{(1)} \leq w_l^{(t)} \leq w_l^{(T)} \quad l = 1, ..., L;\ t = 2, ..., T-1\,, \tag{13}$$

the initial ($k = 0$) population of $P$ individuals/trial solutions is selected to have a set of words with the maximally-diversified chromosomal content. Towards this end, the procedure described in [34] has been adapted to integer-variables, while a further correlation check on each couple of generated words has been performed for having more/different schemata in the initial population. Successively ($k > 1$), standard *GA* operators, namely the *roulette-wheel* selection, the *single-point crossover*, and the *mutation* [46], properly customized to deal with integer-coded chromosomes, are iteratively applied - until the convergence - to the initial trial solutions to generate new populations. More in detail, for each new trial solution/offspring, the *HF* values associated to the corresponding tiling word are computed and the condition (9) is verified, otherwise the *IGA* operators are re-applied until a new ($k \rightarrow k+1$) admissible word is obtained. To preserve the best individual during the iterative process, the new population undergoes elitism [46]. The *IGA* optimization loop is iterated until the generation of a tiling solution that completely fulfills the mask constraints $U(u, v)$ [i.e., $\chi\left(\mathbf{c}^{opt}, \boldsymbol{\alpha}^{opt}, \boldsymbol{\beta}^{opt}\right) = 0$] or when the maximum number of iterations $K$ has been carried out or when the *GA*-stagnation condition

$$\frac{\left|K_{st}\chi_k^{(best)} - \sum_{j=1}^{K_{st}}\chi_j^{(best)}\right|}{\chi_k^{(best)}} \leq \gamma_{st} \tag{14}$$

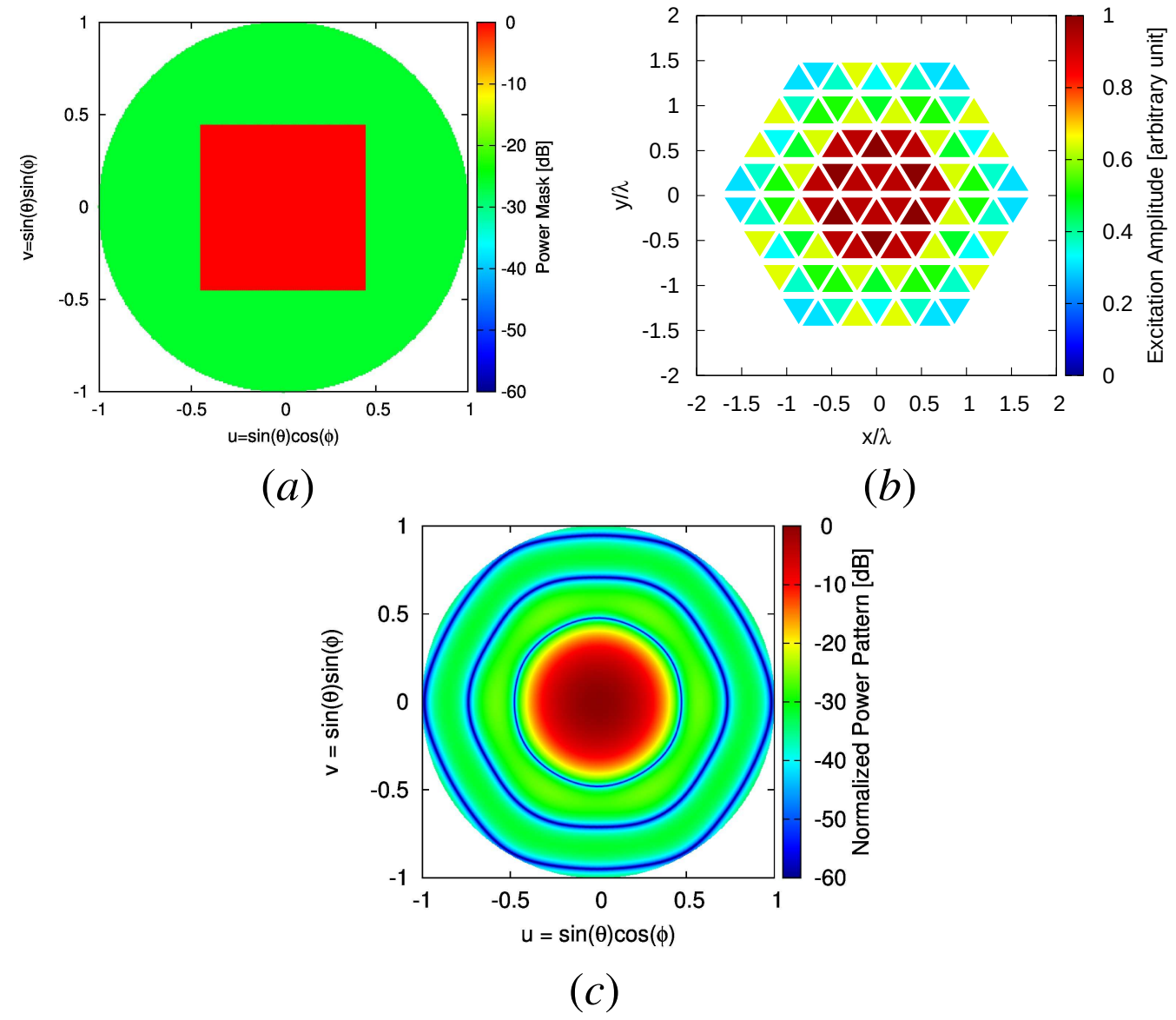

Figure 7. *Test Case 1* ($N_\ell = 4$, $N = 96$, $\rho = \frac{\sqrt{3}}{4}\lambda$, $T = 2.32848 \times 10^5$) - Plot of $(a)$ the power mask, $U(u,v)$, $(b)$ the distribution of the reference amplitudes, $\boldsymbol{\alpha}^{ref}$, and $(c)$ the reference normalized power pattern, $\left|\mathbf{E}^{ref}(u,v)\right|^2$.

arises ($k = K_{conv}$, $k > K_{st}$). In (14), $\chi_k^{(best)} = min_{p=1,..,P}\left\{\chi_k^{(p)}\right\}$, while $K_{st}$ and $\gamma_{st}$ are a user-defined number of iterations and a fixed numerical threshold, respectively.

## IV. Numerical Results and Comparative Assessment

The behavior of the design methodologies presented in the previous section is here analyzed. The *EDM* is first applied to the synthesis of hexagonal arrays with small/medium size apertures. Then, the *CDM* is exploited for synthesizing larger arrays and the *IGA* is compared with the *BGA* [34] to highlight the advantages of the integer coding. A comparative assessment between a diamond-tiled array and a domino-tiled array composing a hexagonal array aperture is carried out and the antenna performance of a selected solution are analyzed by considering a realistic antenna model with mutual coupling, as well.

### *A. EDM for Small/Medium-Array Synthesis*

The first numerical example deals with a hexagonal array with $N_\ell = 4$ and $\rho = \frac{\sqrt{3}}{4}\lambda$ composed by $N = 96$ ideal/isotropic elements (i.e., $\mathbf{p}_{rs}(u,v) = 1, \forall r, \forall s$). The user-defined power mask $U(u,v)$, which mathematically codes the synthesis constraints, is characterized by a sidelobe rejection outside the main lobe of $-26$ [dB], while the mainlobe region has been centered in the visible domain [i.e., the mainlobe peak is expected along broadside, $(u_0, v_0) = (0, 0)$] with an extension of 0.9 along $u$ and $v$ as shown in Fig. 7$(a)$. The reference excitation coefficients, used in (12), have been derived with a convex programming (*CP*) based approach [47] and they are shown in Fig. 7$(b)$ along with the corresponding power pattern [Fig. 7$(c)$], whose pattern features are reported in Tab. III. Only the amplitudes excitations, $\boldsymbol{\alpha}^{ref}$, are shown in Fig. 7 since the phase terms are all zero (i.e., $\boldsymbol{\beta}^{ref} = \mathbf{0}$) because of the broadside steering.

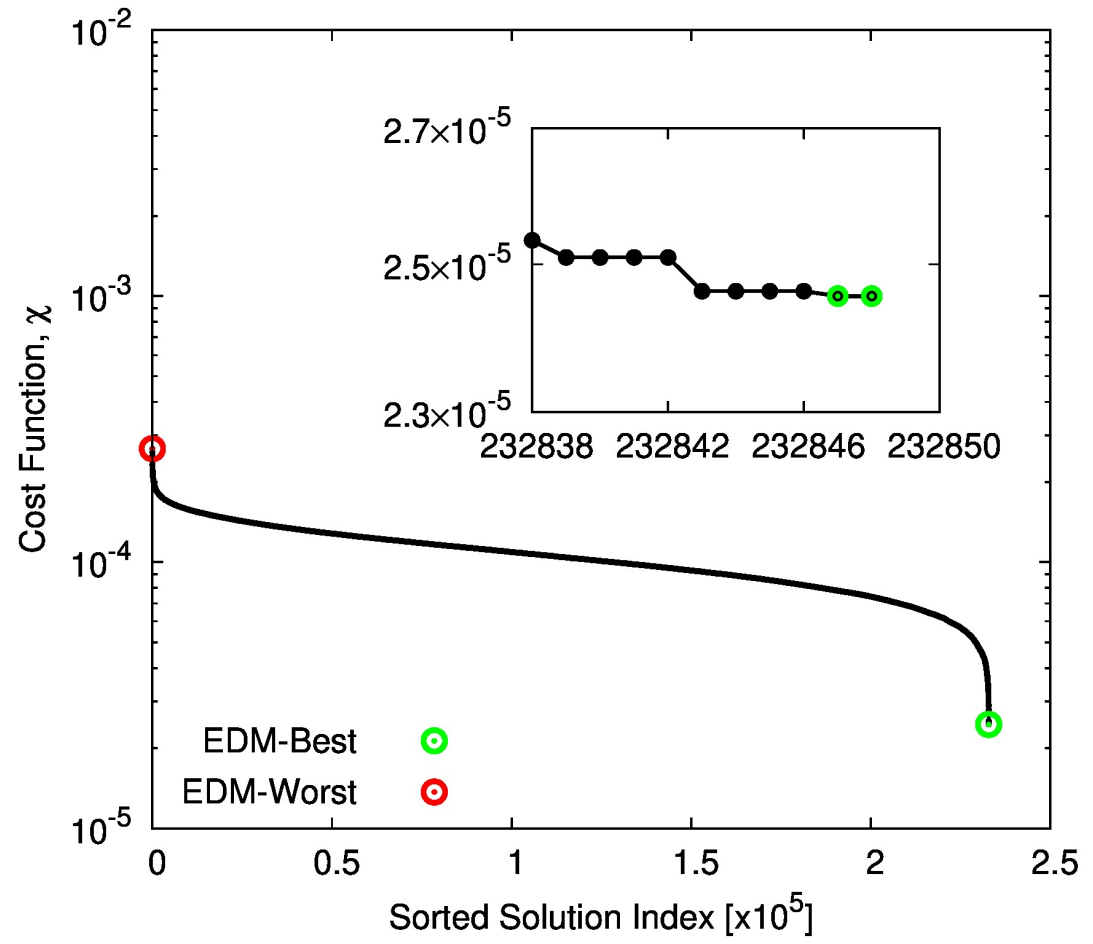

Figure 8. *EDM Synthesis* (*Test Case 1*: $N_\ell = 4$, $N = 96$, $\rho = \frac{\sqrt{3}}{4}\lambda$, $T = 2.32848 \times 10^5$) - Values of the cost function (4) for the whole set of complete tiling configurations, $\{\mathbf{c}^{(t)}; t = 1, ..., T\}$, ordered from the worst tiling, $\mathbf{c}^{worst}$, to the best one, $\mathbf{c}^{opt}$.

In order to generate the whole set of $T = 2.32848 \times 10^5$ tiling words, $\{\mathbf{w}^{(t)}; t = 1, ..., T\}$ and the corresponding sub-array configurations, $\{\mathbf{c}^{(t)}; t = 1, ..., T\}$, to compute the excitation coefficients, $\{\boldsymbol{\alpha}^{(t)}, \boldsymbol{\beta}^{(t)}; t = 1, ..., T\}$, and to evaluate the cost function (4) for each $t$-th trial solution, $\chi^{(t)}$, the *EDM* run for about 6 hours and 40 minutes on a $1.6GHz$ PC with $8GB$ of RAM. The values of the cost function for the complete set of $T$ tiling configurations are sorted in Fig. 8. The "best" and the "worst" cost function values are equal to $\chi^{opt} = 2.46 \times 10^{-5}$ and $\chi^{worst} = 2.67 \times 10^{-4}$, respectively. Those values correspond to the sub-array layouts in Figs. 9$(a)$-9$(b)$ and Fig. 9$(c)$ for the best, $\mathbf{c}^{opt}$, and the worst, $\mathbf{c}^{worst}$ [$\mathbf{c}^{worst} \triangleq \arg\left(\max_{t=1,..,T}\left\{\chi\left(\mathbf{c}^{(t)}\right)\right\}\right)$], solution, respectively. The solutions in Fig. 9$(a)$ and Fig. 9$(b)$ are both optimal since the corresponding sub-array arrangements are equal except for a mirroring with respect to the $x$ axis. Therefore, the radiated power patterns have the same behavior as confirmed by the values of the pattern indexes in Tab. III and illustrated by the power pattern cuts along the principal planes ($\phi = 0$ [deg] and $\phi = 90$ [deg]) shown in Fig. 9$(d)$ and Fig. 9$(e)$, respectively. As it can be observed from Fig. 9$(d)$, the *EDM* pattern slightly violates the mask because of an increment of the secondary lobes of 0.96 [dB] ($SLL^{opt} = -25.04$ [dB] - Tab. III). For the sake of completeness, the sub-array configuration, the amplitude coefficients [Fig. 9$(c)$], and the cuts of the radiated power pattern [Figs. 9$(d)$-$(e)$] of the "worst" tiling solution are reported, as well. In this case, the secondary lobes grow up to $SLL^{worst} = -22.43$ [dB] (Tab. III) with a deterioration of more than 3.5 [dB] with respect to the reference solution and of about 2.6 [dB] as compared to the best tilings.

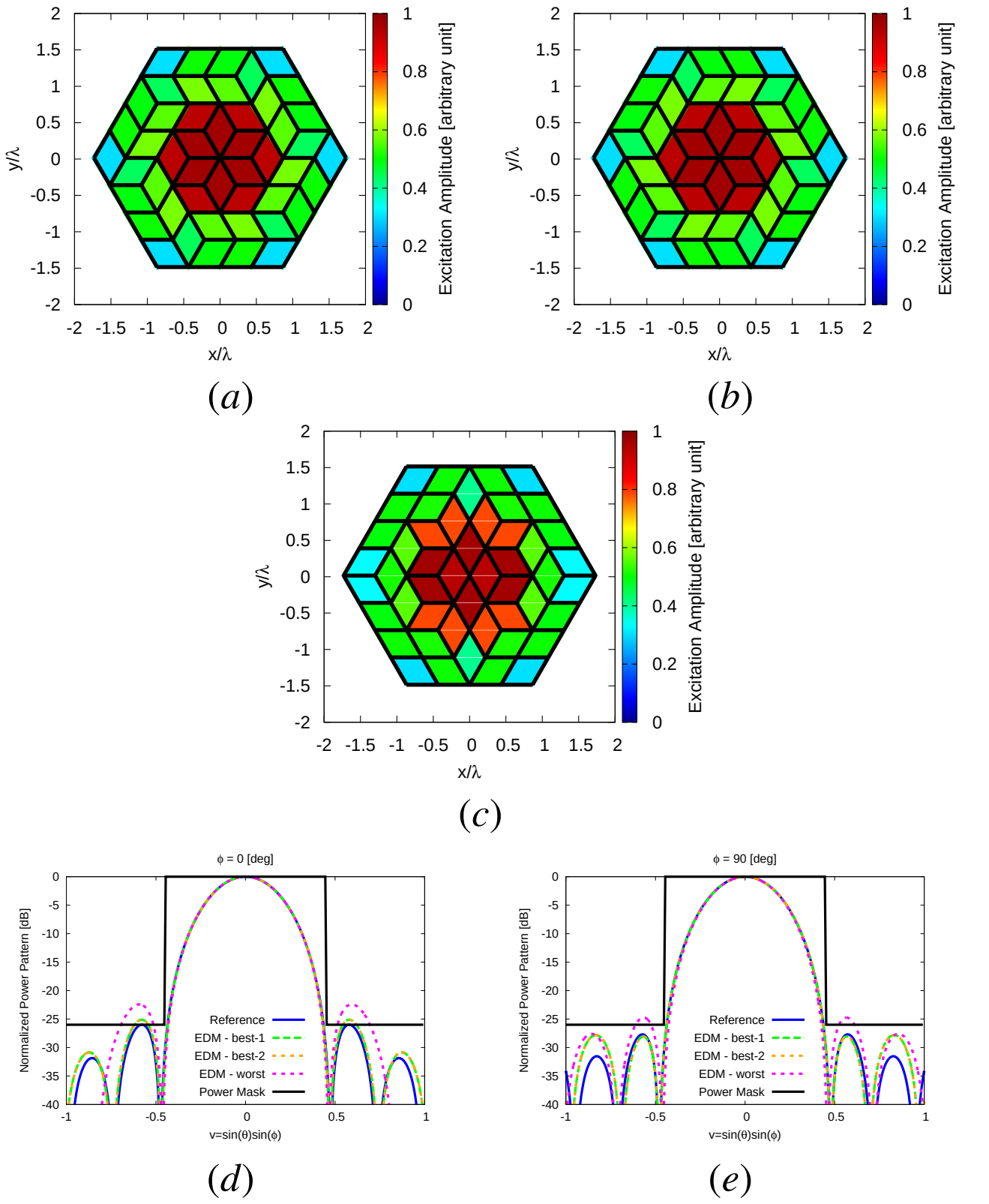

Figure 9. *EDM Synthesis* (*Test Case 1*: $N_\ell = 4$, $N = 96$, $\rho = \frac{\sqrt{3}}{4}\lambda$, $T = 2.32848 \times 10^5$) - Plot of the synthesized sub-array amplitude coefficients for ($a$)($b$) the two optimal/best solutions ($\mathbf{c}^{opt}$, $\boldsymbol{\alpha}^{opt}$) and ($c$) the worst solution ($\mathbf{c}^{worst}$, $\boldsymbol{\alpha}^{worst}$). Power pattern cuts along ($d$) the $\phi = 0$ [deg] and ($e$) the $\phi = 90$ [deg] planes.

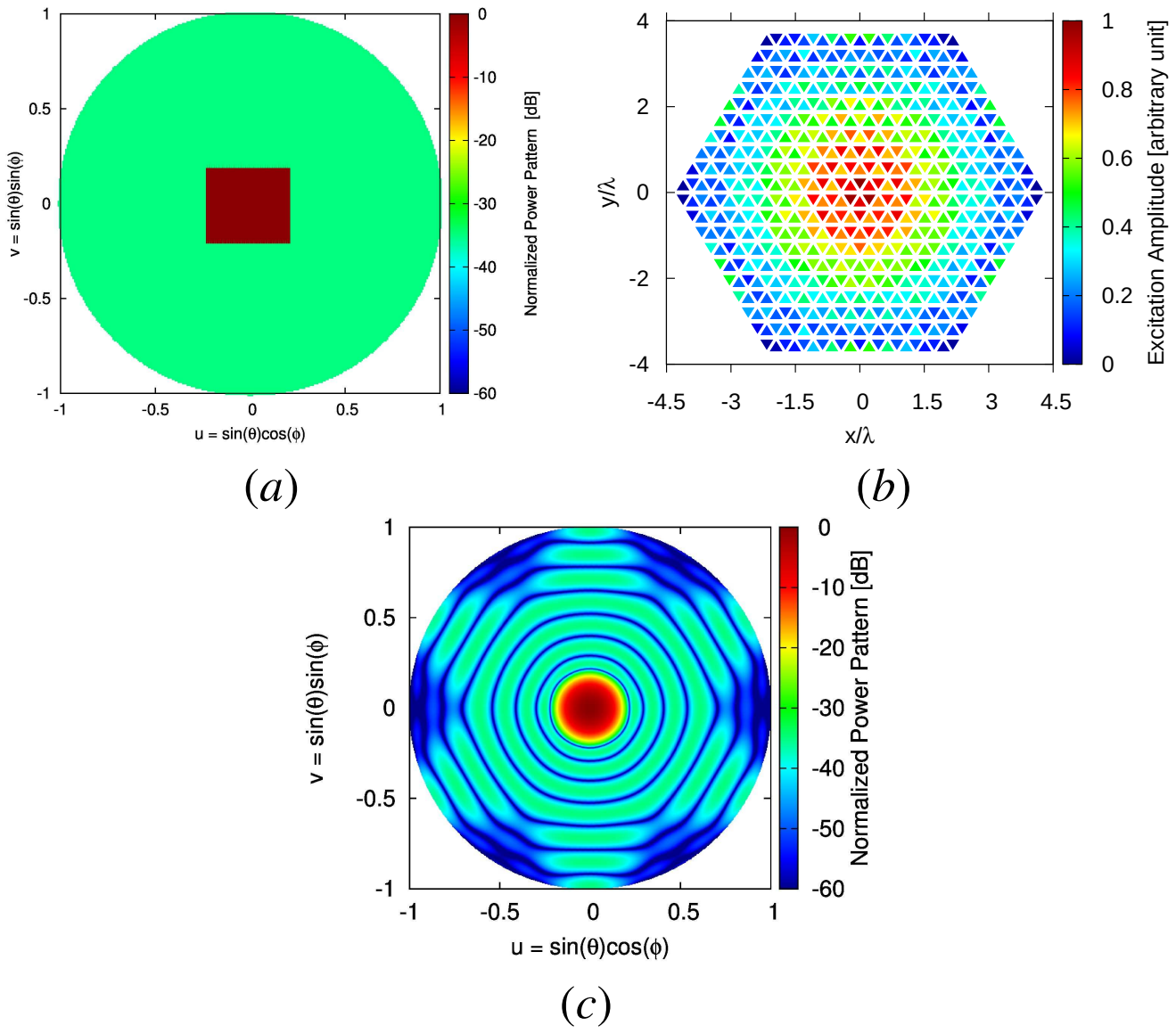

Figure 10. *Test Case 2* [$N_\ell = 10$, $N = 600$, $\rho = \frac{\sqrt{3}}{4}\lambda$, $T \simeq 9.265 \times 10^{33}$; $(u_0, v_0) = (0.0, 0.0)$] - Plot of ($a$) the power mask, $U(u,v)$, ($b$) the distribution of the reference amplitudes, $\boldsymbol{\alpha}^{ref}$, and ($c$) the reference normalized power pattern, $\left|\mathbf{E}^{ref}(u,v)\right|^2$.

Table III
*EDM Synthesis* (*Test Case 1*: $N_\ell = 4$, $N = 96$, $\rho = \frac{\sqrt{3}}{4}\lambda$, $T = 2.32848 \times 10^5$) - Radiation indexes ($SLL$, $D$, $HPBW_{az}$, $HPBW_{el}$) and cost function values ($\chi$).

| | $SLL$ [dB] | $D$ [dBi] | $HPBW_{az}$ [deg] | $HPBW_{el}$ [deg] | $\chi$ [$\times 10^{-5}$] |
|---|---|---|---|---|---|
| $Reference$ | $-26.00$ | 19.42 | 21.10 | 21.07 | $-$ |
| $EDM-best-1$ | $-25.04$ | 19.39 | 21.10 | 21.08 | 2.46 |
| $EDM-best-2$ | $-25.04$ | 19.39 | 21.10 | 21.08 | 2.46 |
| $EDM-worst$ | $-22.43$ | 19.40 | 21.16 | 20.88 | 26.70 |

## B. *CDM for Large-Array Synthesis*

The second example is devoted to the numerical assessment of the *CDM* for the design of larger hexagonal arrays. Towards this aim, a regular array aperture with $N_\ell = 10$, $\rho = \frac{\sqrt{3}}{4}\lambda$, and $N = 600$ elements has been considered. The mask $U(u,v)$ has been chosen, also in this case, with a mainlobe region centered in the visible range, but having a reduced extension of 0.45 along both $u$ and $v$ directions [Fig. 10($a$)] as well as a sidelobe rejection of $-35$ [dB]. The *CP*-synthesized [47] reference excitations of the fully-populated array and the radiated power pattern are given in Fig. 10($b$) and Fig. 10($c$), respectively. Because of the stochastic nature of the *IGA*, a set of 100 *CDM* optimizations has been run by setting the control parameters as follows: $p_c = 0.9$ (crossover probability), $p_m = 0.01$ (mutation probability), $P = 542$ (population size) and $K = 1000$ (maximum number of iterations). More specifically, the values of $p_c$ and $p_m$ have been set according to [46], while $P$ and $K$ have been chosen following the same criterion used in [34] so that $P \times K$ is about 5% of the cardinality of the solution space equal to $T \simeq 9.265 \times 10^{33}$. The cost function of the best individual of the population versus the iteration index $k$ ($k = 0, ..., K$) is shown in Fig. 11($a$) in correspondence with 10 randomly-chosen initialization seeds[1]. As it can be observed, the optimization run labeled as *IGA*-2 reached the lowest value of the cost function ($\chi_{k=K}^{(IGA-2)} \simeq 4.44 \times 10^{-6}$ - Tab. IV). The corresponding sub-array layout is reported in Fig. 12($a$) along with the radiated power pattern [Fig. 12($b$)]. For the sake of clarity, the *CDM* power pattern cuts are compared in Fig. 12($c$) ($\phi = 0$ [deg]) and Fig. 12($d$) ($\phi = 90$ [deg]) with the reference ones, while the corresponding radiation features are given in Tab. IV. It is worth pointing out that, despite the clustering of the radiating elements, the *SLL* of the tiled solutions turns out to be only 1.26 [dB] above the peak sidelobe level of the reference solution.

To assess the advantages of exploiting a customized *IGA*, the same set (i.e., using the same initial populations of trial solutions of the *IGA*) of 100 optimizations has been run with the *BGA* by setting the parameters of the evolutionary operators as in [34]. The behavior of the cost function of the *IGA* and the *BGA* simulations yielding the lowest $\chi$ values is compared in Fig. 11($b$). As one can notice, both methods converge to very small values of the cost function (i.e., $\chi_{k=K}^{(IGA)} \simeq 4.44 \times 10^{-6}$ vs. $\chi_{k=K}^{(BGA)} \simeq 4.51 \times 10^{-6}$ - Tab. IV), even though the *IGA* slightly outperforms the *BGA*. The closeness of both tiling solutions to the pattern-mask is visually confirmed by the plots of the power pattern cuts in Figs. 12($c$)-($d$) where the *IGA* and *BGA* curves essentially overlap. However, it is worth highlighting that the number of iterations to reach the convergence, $k_{conv}$, is almost half for the *IGA*. Indeed, $k_{conv}^{(IGA)} = 186$ vs $k_{conv}^{(BGA)} = 370$ [Fig. 11($b$)] with a reduction of the *CPU*-time from $\tau^{(BGA)} = 286$ [sec] down to $\tau^{(IGA)} = 50$ [sec] (Tab. IV), which corresponds to a computational saving of about 83 %.

[1]To keep a clear visualization, but without altering the meaning of the results also guaranteed by the random choice of the 10 samples, no more curves have been added to the plot.

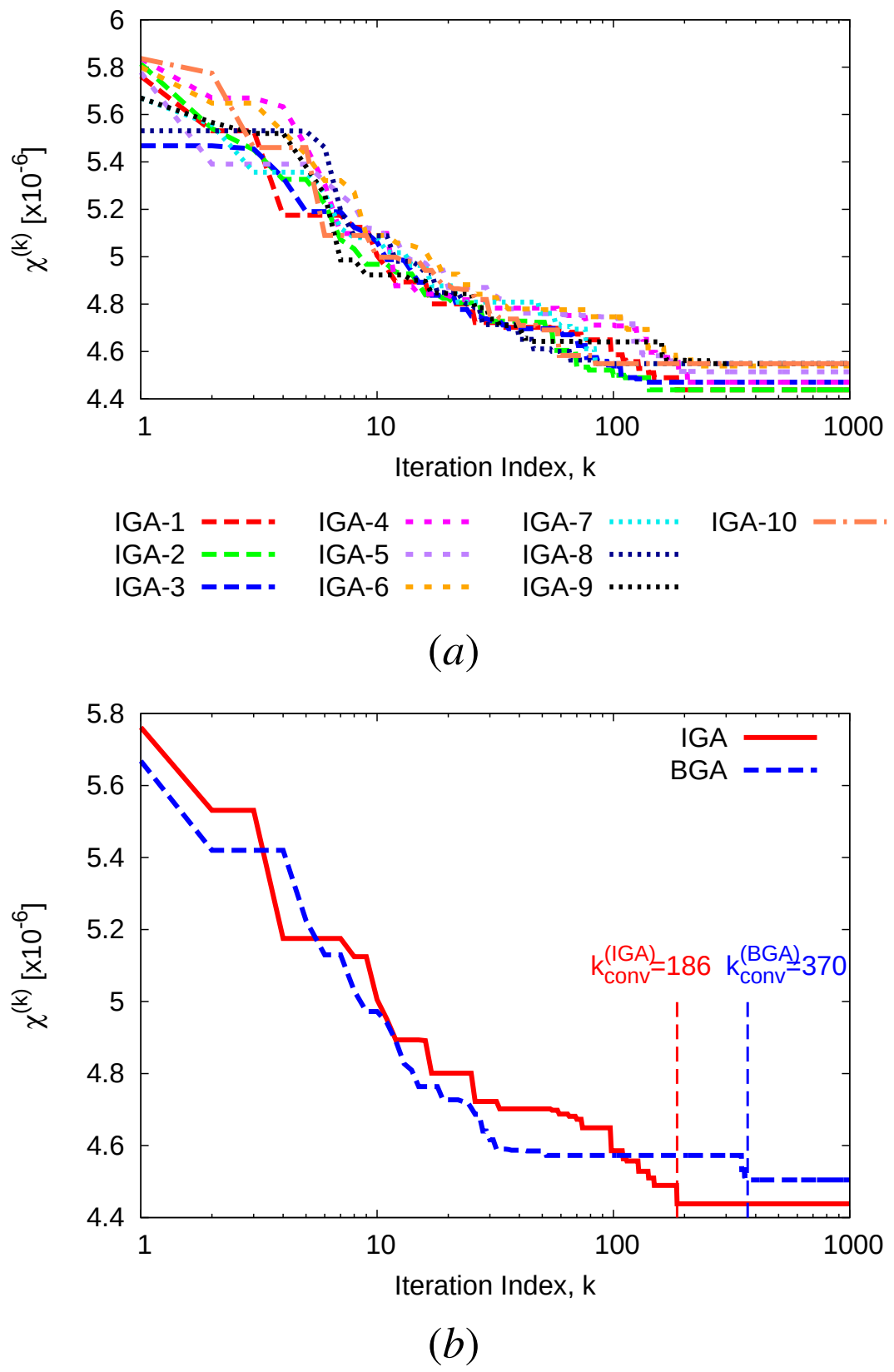

Figure 11. *CDM Synthesis* [*Test Case 2* - $N_\ell = 10$, $N = 600$, $\rho = \frac{\sqrt{3}}{4}\lambda$, $T \simeq 9.265 \times 10^{33}$; $(u_0, v_0) = (0.0, 0.0)$] - Behavior of the optimal value of the cost function (4) versus the iteration index, $k$, ($a$) for 10 randomly-chosen initialization seeds and ($b$) for the best simulations among 100 sample runs when using the *IGA* and the *BGA*.

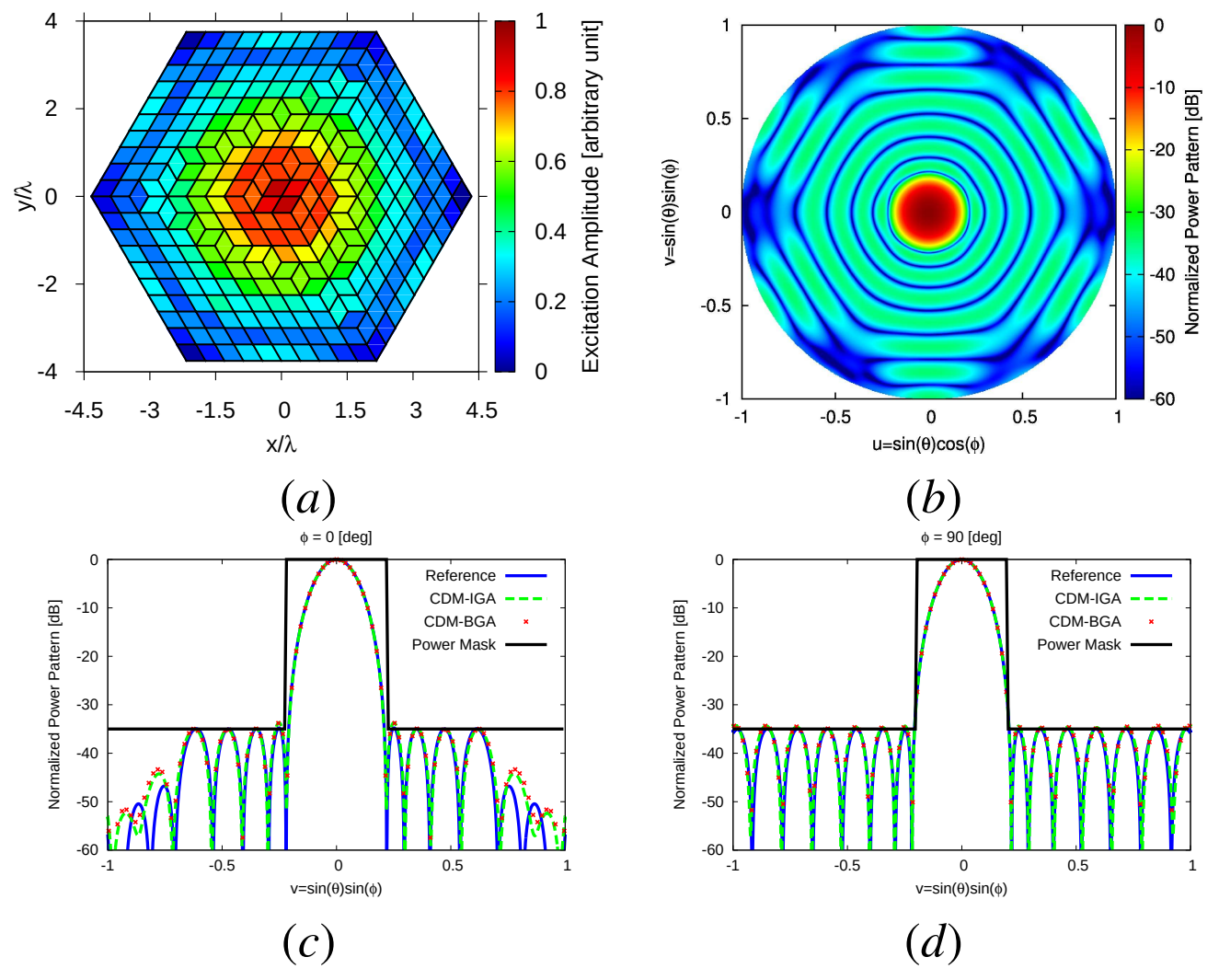

Figure 12. *CDM Synthesis* [*Test Case 2* - $N_\ell = 10$, $N = 600$, $\rho = \frac{\sqrt{3}}{4}\lambda$, $T \simeq 9.265 \times 10^{33}$; $(u_0, v_0) = (0.0, 0.0)$] - Plot of ($a$) the sub-array amplitude coefficients and the corresponding normalized power pattern: ($b$) 2D color map and cuts along ($c$) the $\phi = 0$ [deg] and ($d$) the $\phi = 90$ [deg].

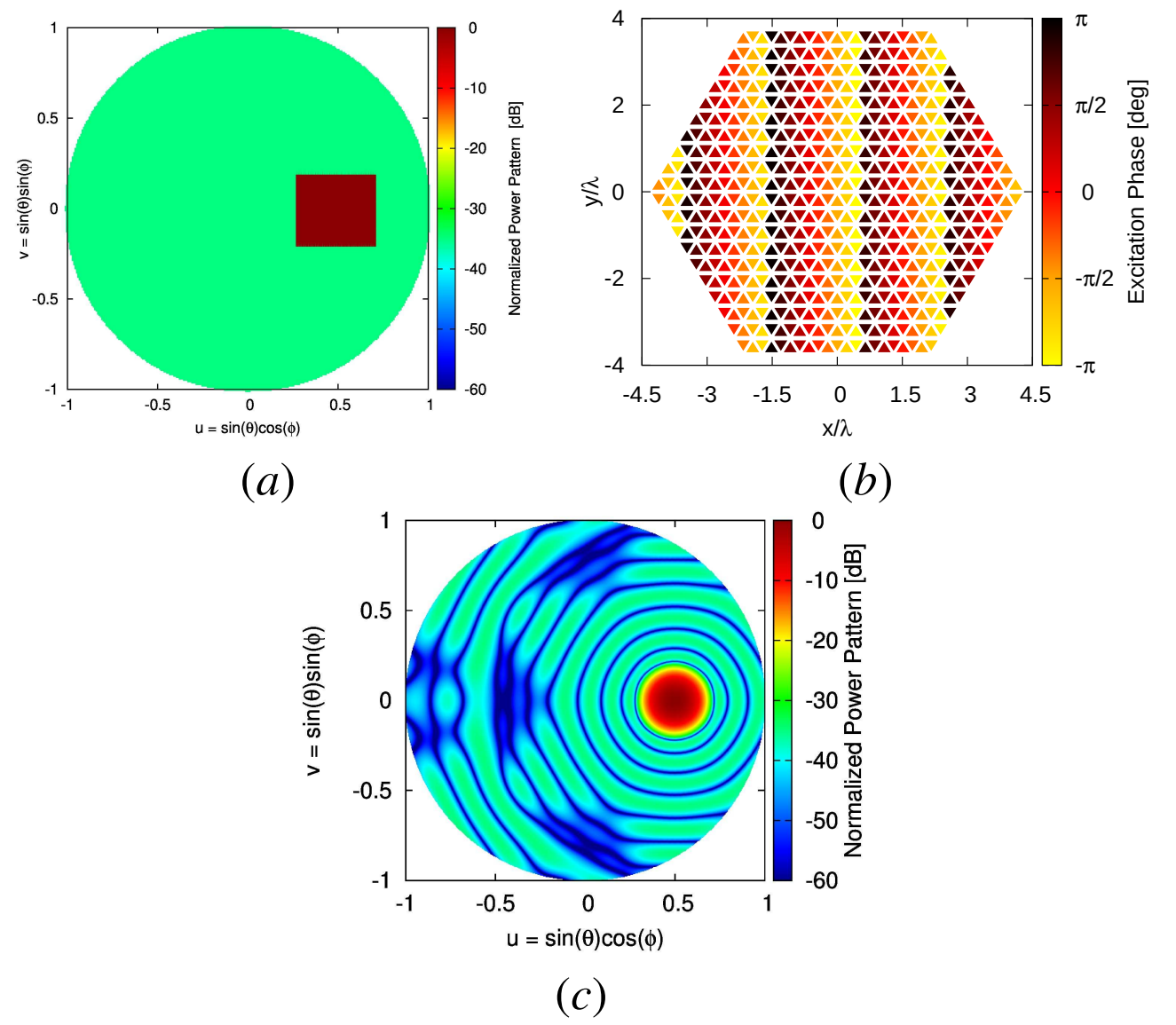

Figure 13. *Test Case 3* [$N_\ell = 10$, $N = 600$, $\rho = \frac{\sqrt{3}}{4}\lambda$, $T \simeq 9.265 \times 10^{33}$; $(u_0, v_0) = (0.5, 0.0)$] - Plot of ($a$) the power mask $U(u, v)$, ($b$) the distribution of the reference phase coefficients, $\boldsymbol{\beta}^{ref}$, and ($c$) the reference power pattern, $\left|\mathbf{E}^{ref}(u, v)\right|^2$.

In the third example, the array aperture and the sidelobe suppression level of the previous case have been kept, but the pattern beam has been constrained to point along the direction $(u_0, v_0) = (0.5, 0.0)$ as indicated by the mask $U(u, v)$ in Fig. 13($a$). Once *CP*-computed the amplitude [Fig. 10($b$)] and the phase [Fig. 13($b$)] distributions of the reference fully-populated array, whose power pattern is given in Fig. 13($c$), the *CDM* has been applied to synthesize the tiled array. The best result among 100 different *IGA*-based optimizations is shown in Fig. 14. As expected, the layout of the diamond tiles within the aperture turns out to be different from that of the broadside-steered case [Fig. 14($a$) and Fig. 14($b$) vs. Fig. 12($a$)]. Concerning the power pattern [Fig. 14($c$)], it faithfully matches the reference one even though there is an increment of the sidelobe level ($SLL^{IGA} = -31.57$ [dB] vs. $SLL^{ref} = -35.00$ [dB]) with respect to the second test case (Tab. IV) as also confirmed by the higher value of the cost function at the convergence (i.e., $\left.\chi_{k=K}^{(IGA)}\right|_{(u_0,v_0)=(0.5,0.0)} = 1.60 \times 10^{-5}$ vs. $\left.\chi_{k=K}^{(IGA)}\right|_{(u_0,v_0)=(0.0,0.0)} = 4.44 \times 10^{-6}$).

## C. Comparison with Domino-Tiled Array

The fourth example is devoted to the comparison with the domino-tiling architecture discussed in [34] where the radi-ating elements are positioned on a square grid and clustered into dominoes. Towards this end, a regular ($\ell = 6\lambda$) hexagonal aperture $A$ has been partitioned with equilateral triangular cells having $\rho = 0.6\lambda$ ($\rightarrow N_\ell = 10$) yielding an array of $N^{(hex)} = 600$ elements. The same aperture $A$ has been also covered with domino tiles arranged on a square lattice with inter-element spacing equal to $d_x^{(dom)} = d_y^{(dom)} = 0.4\lambda$. Under these conditions, the domino-tiled array turned out to be composed by $N^{(dom)} = 576$ elements so that $N^{(dom)} \approx N^{(hex)}$. As for the pattern mask $U(u, v)$, the size of the mainlobe region has been reduced to $0.35$ along both $u$ and $v$ directions with respect to the previous test case.

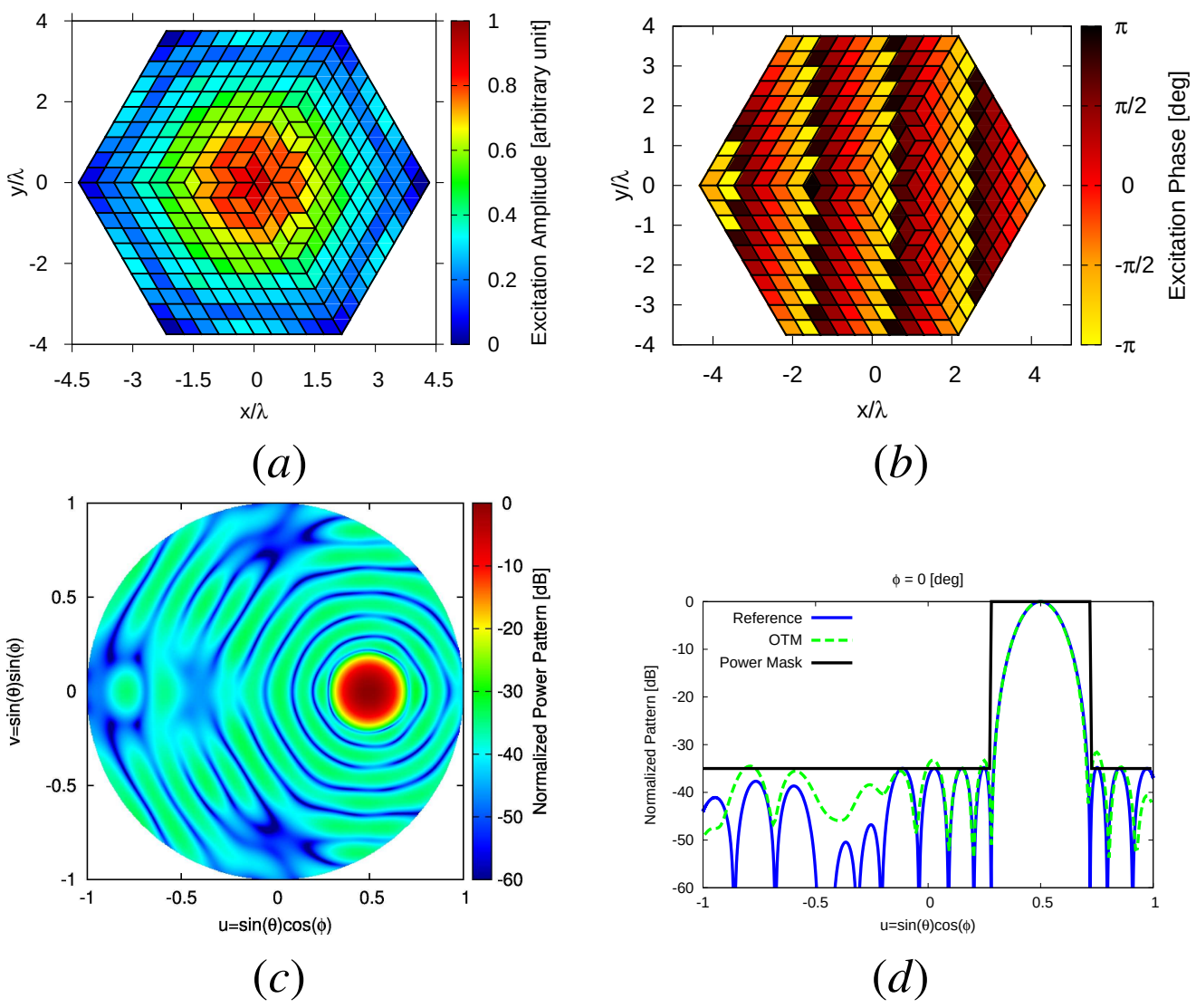

Figure 14. *CDM Synthesis* [*Test Case 3*: $N_\ell = 10$, $N = 600$, $\rho = \frac{\sqrt{3}}{4}\lambda$, $T \simeq 9.265 \times 10^{33}$; $(u_0, v_0) = (0.5, 0.0)$] - Plot of the sub-array (*a*) amplitude and (*b*) phase coefficients, and the radiated normalized power pattern: (*c*) 2D color map and (*d*) cut along the $\phi = 0$ [deg] plane.

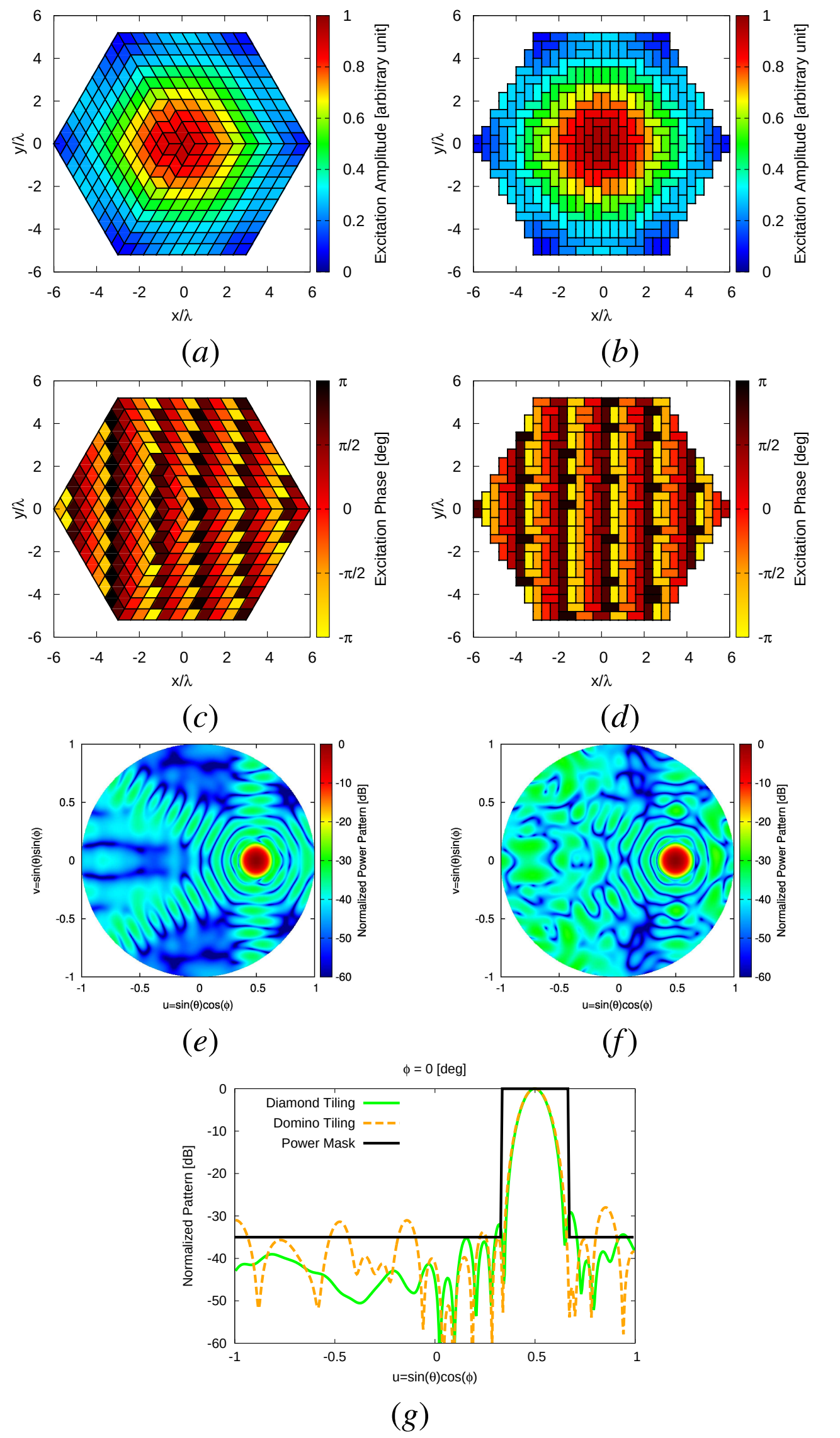

Figure 15. *CDM Synthesis* - Plot of the sub-array (*a*)(*b*) amplitude and (*c*)(*d*) phase coefficients, the normalized power patterns - (*e*)(*f*) 2D color map and (*g*) cut along the $\phi = 0$ plane - for the optimal *CDM* solutions when clustering the array aperture with (*a*)(*c*)(*e*) diamond-shaped tiles [*Test Case 3*: $N_\ell = 10$, $N^{(hex)} = 600$, $\rho = 0.6\lambda$, $T \simeq 9.265 \times 10^{33}$; $(u_0, v_0) = (0.5, 0.0)$] and (*b*)(*d*)(*f*) domino-shaped tiles [*Test Case 3*: $N^{(dom)} = 576$, $d_x^{(dom)} = d_y^{(dom)} = 0.4\lambda$; $(u_0, v_0) = (0.5, 0.0)$].

By applying the *CDM-IGA* and [34]-*BGA* for designing the diamond-tiled and the domino-tiled array, respectively, the arising optimal tiling arrangements and the synthesized distributions of the sub-array amplitudes [Figs. 15(*a*)-(*b*)] and phases [Figs. 15(*c*)-(*d*)] are shown in Fig. 15 together with the corresponding power patterns [Figs. 15(*e*)-(*f*)]. As expected, the domino-based solution [Fig. 15(*f*)] presents secondary lobes higher than those from the diamond partitioning of the aperture [Fig. 15(*e*)] because of the worse approximation of the reference excitations. Indeed, the values of the cost function at convergence are $\chi^{(hex)} = 1.13 \times 10^{-5}$ and $\chi^{(dom)} = 1.20 \times 10^{-4}$. Such an outcome is further highlighted by the cut of the power pattern along the $\phi = 0$ [deg] plane in Fig. 15(*g*) as well as by the values of the sidelobe level (Tab. V). More specifically, it turns out that $SLL^{(dom)} = -27.20$ [dB], that is 2 [dB] above the one of the diamond arrangement ($SLL^{(hex)} = -29.16$ [dB]).

The next experiment has been performed to assess the performance of the optimized tiled-arrays while scanning. By choosing the sub-array configuration and the amplitude in Fig. 15(*a*) and Fig. 15(*b*) as representative examples for the diamond-tiled and the domino-tiled solution, respectively, and by setting the sub-array phases, $\{\beta_q;\ q = 1, ..., Q\}$, by means of (12) with the reference phases set to

$$\beta_{rs}^{ref} = -k\left[x_r \sin\left(\theta_0 + \theta_\gamma\right)\cos\left(\phi_0 + \phi_\gamma\right) + y_s \sin\left(\theta_0 + \theta_\gamma\right)\sin\left(\phi_0 + \phi_\gamma\right)\right] \quad (15)$$

($r = -\left(\frac{N^{(s)}-1}{2}\right), ..., \left(\frac{N^{(s)}-1}{2}\right)$; $s = -N_\ell, ..., N_\ell$; $s \neq 0$), the behavior of the *SLL* and of the directivity, *D*, has been analyzed when scanning the beam around the pointing direction $(\theta_0, \phi_0) = (30, 0)$ [deg] [i.e., $(u_0, v_0) = (0.5, 0.0)$] and within the cone defined by the following angular ranges: $0 \leq \phi_\gamma < 360$ [deg] and $-30 \leq \theta_\gamma < 30$ [deg]. With reference to the polar color maps in Fig. 16, it turns out that the domino tiling shows good performance only along the $\phi = 0$ [deg] plane [Figs. 16(*b*)-16(*d*)], while the diamond clustering affords good values of the pattern features along three planes spaced of 60 [deg] along the $\phi$ direction [Figs. 16($a$)-16($c$)]. For instance, let us assume the maximum scan angle be equal to $\theta_\gamma = 15$ [deg] and let us evaluate the *SLL* and $D$ over the whole scanning cone ($0 \leq \phi_\gamma < 360\,[deg]$). The peak sidelobe level of the diamond-tiled array is $SLL^{(hex)} = -23.91$ [dB], while $SLL^{(dom)} = -11.68$ [dB] for the domino-tiled one with a worsening of more than 12 [dB]. Moreover, the directivity values turn out to be $D^{(hex)} > 27.73$ [dBi] and $D^{(dom)} > 25.57$ [dBi] within the same scanning cone.

Table IV
*CDM Synthesis* [*Test Case 2* - $N_\ell = 10$, $N = 600$, $\rho = \frac{\sqrt{3}}{4}\lambda$, $T \simeq 9.265 \times 10^{33}$] - RADIATION INDEXES ($SLL$, $D$, $HPBW_{az}$, $HPBW_{el}$), COST FUNCTION VALUES ($\chi$), AND COMPUTATIONAL TIME ($\tau$) WHEN STEERING THE BEAM ALONG $(u_0, v_0) = (0.0, 0.0)$ AND $(u_0, v_0) = (0.5, 0.0)$.

| | $SLL$ [dB] | $D$ [dBi] | $HPBW_{az}$ [deg] | $HPBW_{el}$ [deg] | $\chi$ - | $\tau$ [sec] |
|---|---|---|---|---|---|---|
| | *Broadside Mainlobe* - $(u_0, v_0) = (0.0, 0.0)$ | | | | | |
| $Reference$ | $-35.00$ | 26.79 | 9.00 | 8.91 | $-$ | $-$ |
| $IGA$ | $-33.74$ | 26.77 | 9.00 | 8.93 | $4.44 \times 10^{-6}$ | 50 |
| $BGA$ | $-33.71$ | 26.77 | 9.00 | 8.93 | $4.51 \times 10^{-6}$ | 286 |
| | *Steered Mainlobe* - $(u_0, v_0) = (0.5, 0.0)$ | | | | | |
| $Reference$ | $-35.00$ | 26.13 | 10.42 | $-$ | $-$ | $-$ |
| $IGA$ | $-31.57$ | 26.12 | 10.36 | $-$ | $1.60 \times 10^{-5}$ | 48 |

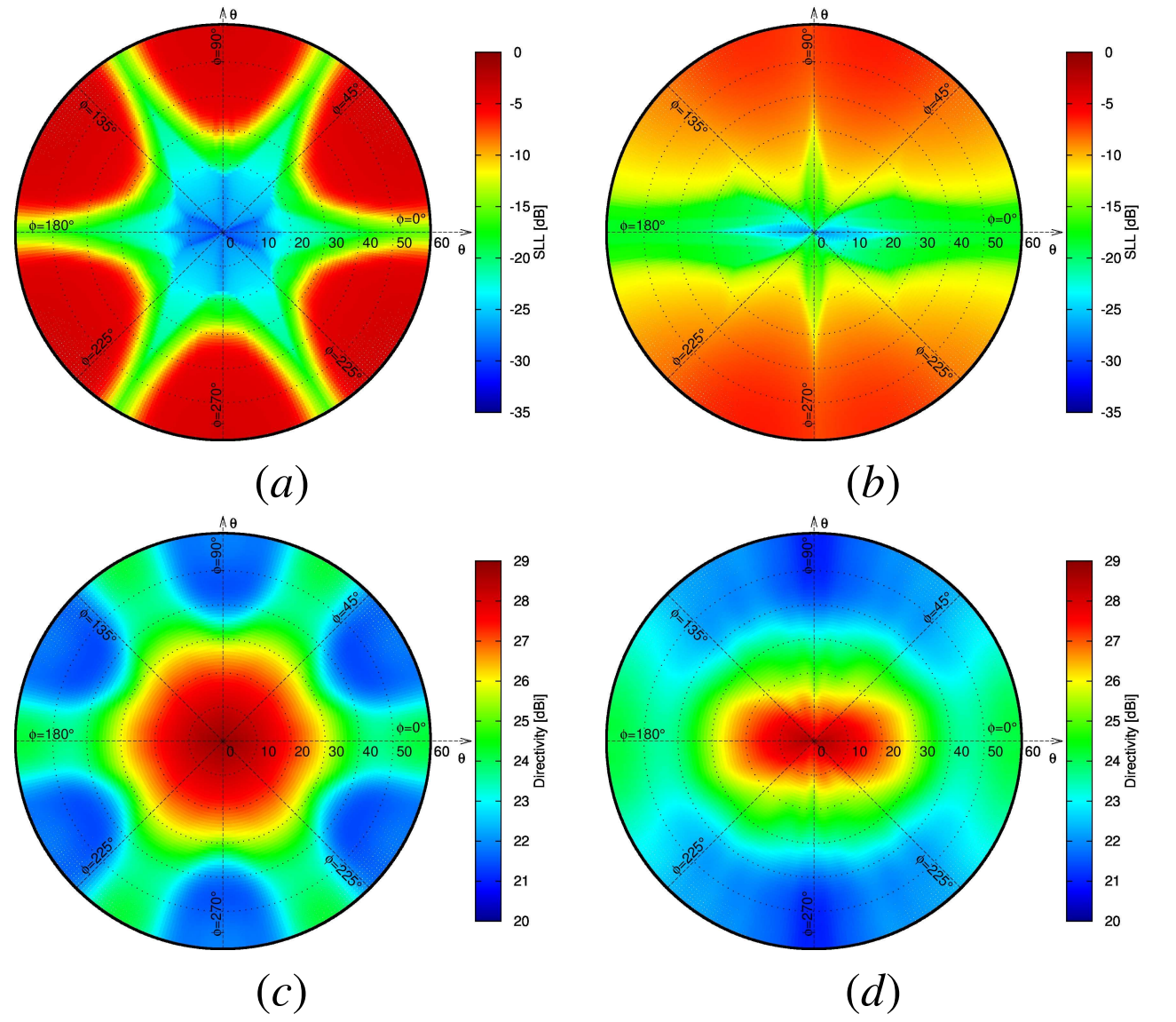

Figure 16. *CDM Synthesis* [*Test Case 3*: $N_\ell = 10$, $N = 600$, $\rho = \frac{\sqrt{3}}{4}\lambda$, $T \simeq 9.265 \times 10^{33}$; $(u_0, v_0) = (0.5, 0.0)$] - Plot of ($a$)($b$) the *SLL* and ($c$)($d$) the $D$ values of the patterns generated by the tiling configurations in ($a$)($c$) Fig. 15($a$) and Fig. 15($c$) and ($b$)($d$) Fig. 15($b$) and Fig. 15($d$) when scanning the mainlobe around the pointing direction $(\theta_0, \phi_0) = (30, 0)$ [deg] within the cone $\{-30 \leq \theta_\gamma < 30$ [deg]; $0 \leq \phi_\gamma < 360$ [deg]$\}$.

Finally, the performance of the optimized diamond-tiling in Figs. 15($a$)-15($c$) have been validated against realistic radiating elements [i.e., $\mathbf{p}_{rs}(u, v) \neq 1$], including mutual coupling effects, as well. Towards this end, a coaxial-fed patch antenna resonating at $f = 7.9$ [GHz] has been used as elementary radiator, while the entire array structure (i.e., the dielectric substrate, the ground plane, the $N = 600$ elements array together with the metallic coaxial connector) has been modelled with CST Microwave Studio [Fig. 17($a$)]. The full-wave power pattern has been generated when pointing the beam at $(u_0, v_0) = (0.5, 0.0)$, as in Fig. 15, as well as along the direction $(u_0, v_0) = (0.0, 0.5)$ and in broadside $(u_0, v_0) = (0.0, 0.0)$. From the comparison of the realistic pattern with the ideal one (i.e., isotropic sources array)[2] along the $\phi = 0$ [deg] cut [Figs. 17($b$)-17($d$)] and the $\phi = 90$ [deg] cut [Figs. 17($c$)-17($e$)], one can infer that the two patterns are very similar with only some minor and almost negligible deviations towards the boundaries of the visible range ($u = \pm 1$, $v = \pm 1$), thus highlighting the facts that the beam of the patch antenna is broad and the mutual coupling effects are not significant in the considered array structure [Fig. 17($a$)].

## V. CONCLUSIONS AND REMARKS

The modular synthesis of hexagonal arrays with elements disposed on a regular honeycomb lattice and clustered in diamond-shaped tiles has been addressed. Tiling theorems have been exploited to assess the perfect covering of the array aperture by means of diamond tiles as well as to *a-priori* determine the cardinality of the solution space of the full-aperture tiling configurations. Starting from such a theoretical basis, two tiling strategies have been presented to synthesize array solutions fully covering the available antenna aperture. The former is based on an enumerative approach, while the other relies on a customized version of the *IGA*.

The key methodological advancements of the proposed research work with respect to the state-of-the-art literature can be summarized in the following:

- the modular design of phased array antennas with regular hexagonal apertures through diamond-shaped tiles;
- the exploitation of mathematical theorems and algorithms to define a theoretical framework for the clustering of hexagonal phased arrays;
- the formulation of the synthesis of hexagonal phased arrays clustered with diamond tiles as a mask-constrained power pattern one;
- the integer coding of the tiling words within the schemata driven optimization and the definition of a customized integer-coded *GA*-based tiling approach.

A set of numerical examples, including both ideal-isotropic sources and real-directive antenna elements, has been reported to assess the capabilities and the effectiveness of the proposed

[2] Each pattern has been normalized to its maximum in order to better compare the patterns shape.

Table V
*CDM Synthesis* - RADIATION INDEXES ($SLL$, $D$, $HPBW_{az}$), COST FUNCTION VALUES ($\chi$), AND COMPUTATIONAL TIME ($\tau$) WHEN CLUSTERING THE ARRAY APERTURE WITH DIAMOND-SHAPED TILES [*Test Case 3*: $N_\ell = 10$, $N^{(hex)} = 600$, $\rho = 0.6\lambda$, $T \simeq 9.265 \times 10^{33}$; $(u_0, v_0) = (0.5, 0.0)$] AND $(b)(d)(f)$ DOMINO-SHAPED TILES [*Test Case 3*: $N^{(dom)} = 576$, $d_x^{(dom)} = d_y^{(dom)} = 0.4\lambda$; $(u_0, v_0) = (0.5, 0.0)$].

| | | $SLL$ [dB] | $D$ [dBi] | $HPBW_{az}$ [deg] | $\chi$ - | $\tau$ [sec] |
|---|---|---|---|---|---|---|
| | | *Domino Tiles* | | | | |
| $Reference$ | $Isotropic$ | $-35.00$ | 29.52 | 6.53 | $-$ | $-$ |
| $CDM$ | $Isotropic$ | $-27.20$ | 28.62 | 7.56 | $1.20 \times 10^{-4}$ | 806 |
| | | *Diamond Tiles* | | | | |
| $Reference$ | $Isotropic$ | $-35.00$ | 29.65 | 6.39 | $-$ | $-$ |
| $CDM$ | $Isotropic$ | $-29.16$ | 28.91 | 7.37 | $1.13 \times 10^{-5}$ | 21 |
| $CDM$ | $Full-Wave$ | $-27.69$ | 28.88 | 7.30 | $3.40 \times 10^{-3}$ | $-$ |

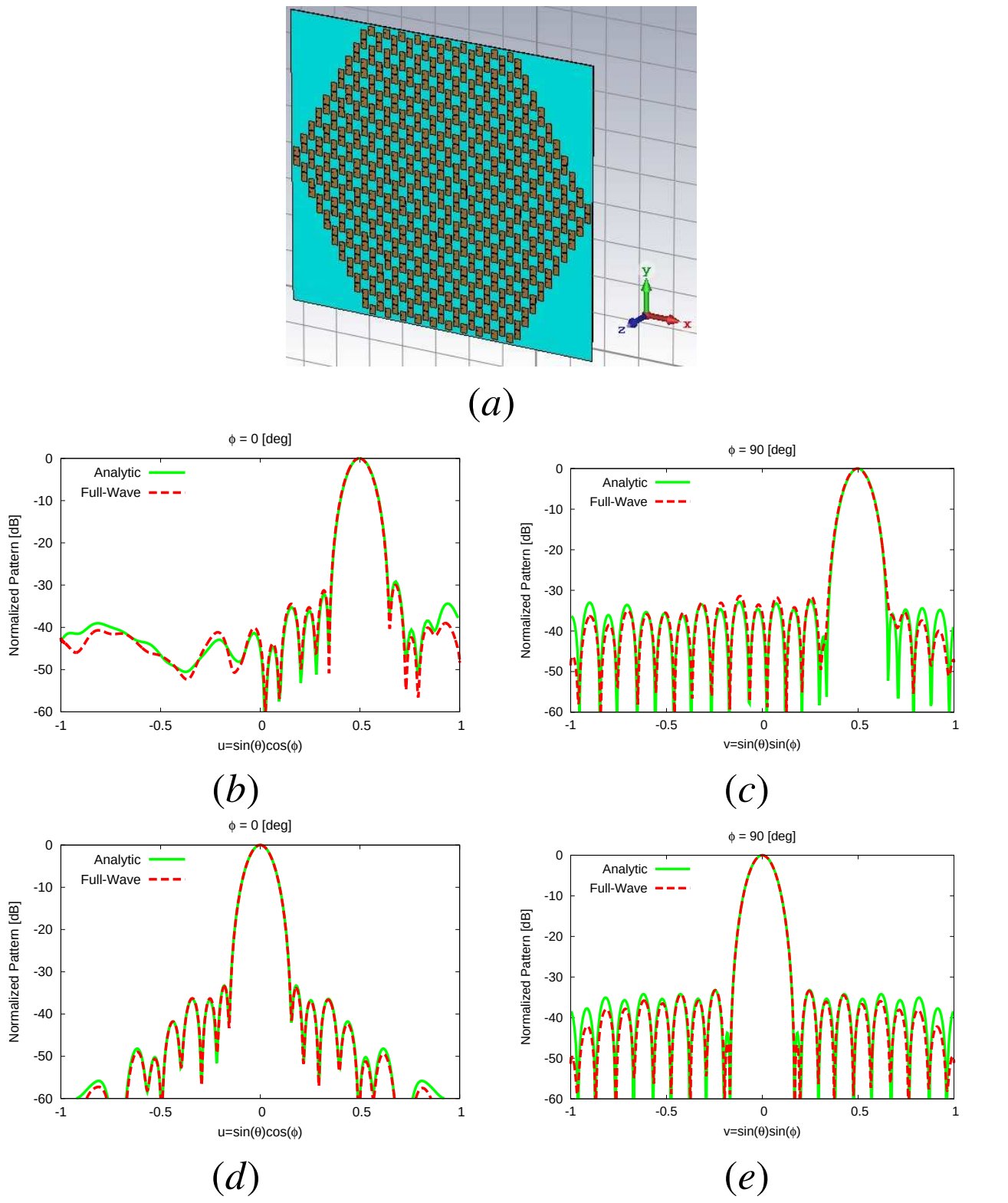

Figure 17. *CDM Synthesis* [*Test Case 3*: $N_\ell = 10$, $N = 600$, $\rho = \frac{\sqrt{3}}{4}\lambda$, $T \simeq 9.265 \times 10^{33}$] - Model of $(a)$ the hexagonal array of coaxial-fed patch antennas and plot of the normalized power pattern cuts along the $(b)(d)$ $\phi = 0$ [deg] plane when $(b)$ $(u_0, v_0) = (0.5, 0.0)$ and $(d)$ $(u_0, v_0) = (0.0, 0.0)$ and along the $(c)(e)$ $\phi = 90$ [deg] plane when $(c)$ $(u_0, v_0) = (0.0, 0.5)$ and $(e)$ $(u_0, v_0) = (0.0, 0.0)$.

synthesis strategies. From the numerical analysis, the following main outcomes can be drawn:

- the *EDM* is efficient in generating all possible tiling configurations and to retrieve the optimal solution, or the multiple optimal solutions in case of the symmetrical tilings, that guarantees the best admissible performance (i.e., the global minimum of the cost function at hand) when the size of the array is small/medium;
- the exploitation of integer-coded chromosomes in the *IGA* implies a non negligible reduction of the computational burden and the improvement of the convergence speed, with respect to the standard *BGA*-based synthesis [34], for designing large arrays;
- the radiation performance (e.g., the sidelobe control and the peak directivity) of an hexagonal aperture partitioned with diamond-shaped tiles are generally better than those from an equivalent array clustered with domino subarrays when scanning the beam over a limited field of view;
- the power pattern generated by an ideal diamond-tiled array is close to that from a realistic one that includes non-isotropic elements and mutual coupling effects.

## *Appendix I* - TILEABILITY THEOREM

An arbitrary hexagonal aperture $A$ having sides of length $l_i$, $i = 1, ..., 6$ is fully coverable with diamond tiles if and only if the opposite sides have the same dimension, namely $l_1 = l_4$, $l_2 = l_5$, and $l_3 = l_6$ [40][41][42]. Such a condition holds true for both regular [Fig. 18$(a)$] and non-regular [Fig. 18$(b)$] hexagons, while the aperture in Fig. 18$(c)$ is un-tileable since $l_1 \neq l_4$, $l_2 \neq l_5$, $l_3 \neq l_6$, and the number of triangles $N$ turns out being odd. As for regular hexagonal apertures ($l_i = \ell$; $i = 1, ..., 6$), as those dealt with in this work, they always satisfy the tileability condition (i.e., $l_1 = l_4$, $l_2 = l_5$, and $l_3 = l_6$) and they can be totally tileable with diamond-shaped tiles.

## *Appendix II* - CARDINALITY THEOREM

By supposing a fully-tileable hexagonal aperture $A$ [Figs. 18$(a)$-$(b)$], the total number $T$ of different tiling configurations that completely cover $A$ is equal to [43]

$$T = \prod_{i=1}^{N_{l_1}} \prod_{j=1}^{N_{l_2}} \prod_{g=1}^{N_{l_3}} \frac{i+j+g-1}{i+j+g-2} \quad (16)$$

where $N_{l_i} \triangleq \frac{l_i}{\rho}$ ($i = \{1, 2, 3\}$) is the number of equilateral triangles (i.e., unit cells) along the $i$-th side of the hexagonal aperture.
A set of cardinality values ($T \leftarrow T^{(hex)}$) in correspondence with different number of elements $N$ is reported in Tab. I. For comparison purposes, the cardinality of the solution space ($T \leftarrow T^{(dom)}$) for the case of rectangular apertures having the same number of elements $N$, but considering domino-shaped tiles as in [34], is reported, as well. As it can be inferred from Tab. I and observed in Fig. 5, $T^{(dom)}$ grows much faster than $T^{(hex)}$ when increasing $N$. This indicates a smaller solution space for hexagonal arrays, thus a higher complexity of the tiling problem at hand whether setting the same constraints and requirements of the domino partitioning - because of the reduced number of admissible solutions - and the need of ad-hoc synthesis techniques.

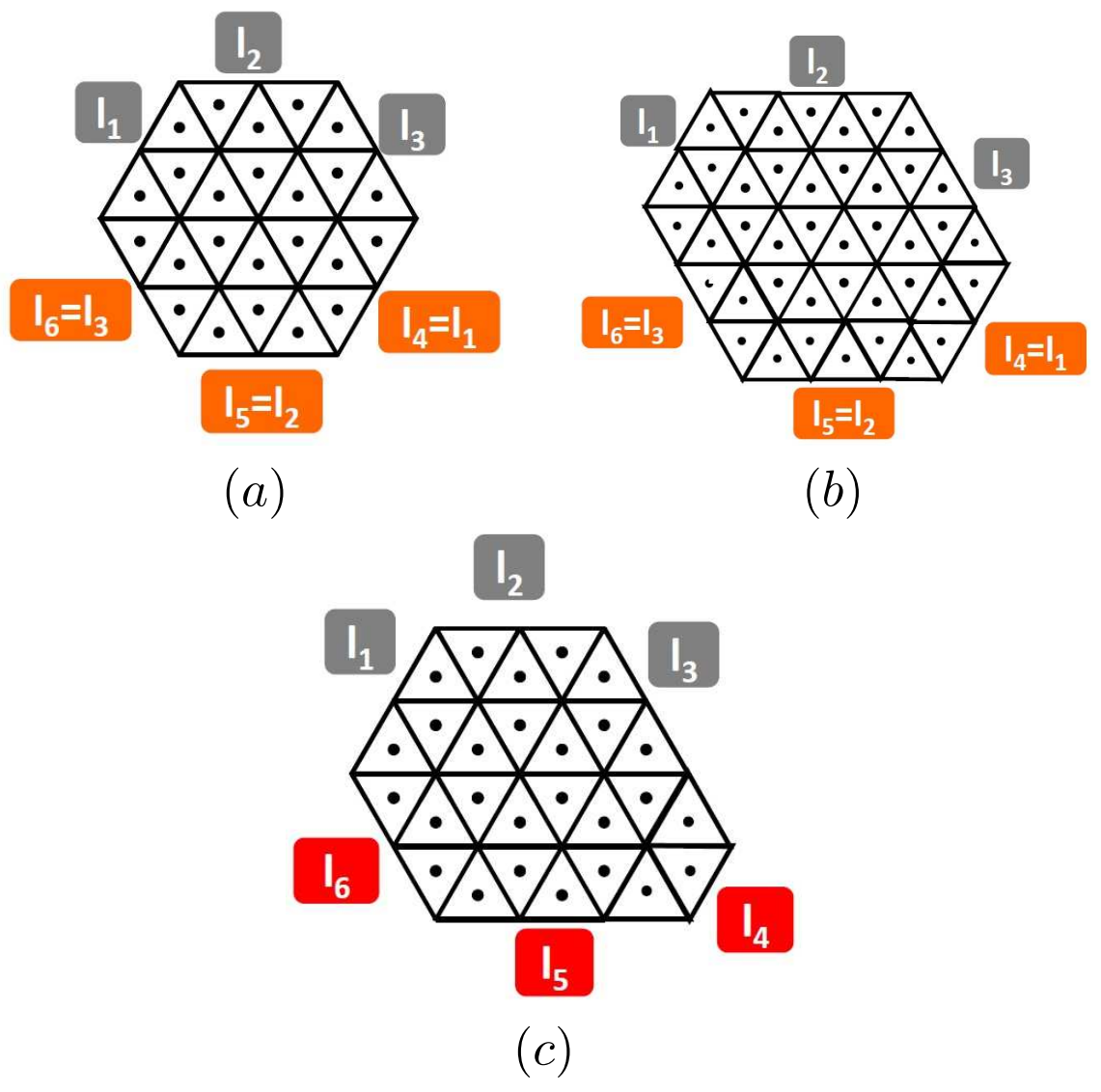

Figure 18. *Illustrative Geometry* - Examples of $(a)$ regular tileable ($l_i = \ell$; $i = 1, ..., 6$), $(b)$ irregular tileable ($l_1 = l_4$, $l_2 = l_5$, $l_3 = l_6$), and $(c)$ irregular non-tileable ($l_1 \neq l_4$, $l_2 \neq l_5$, $l_3 \neq l_6$) hexagonal apertures.

## *Appendix III* - Height Function Computation

To define the height function, let us consider the lattice defined by the sets of vertices $\mathbf{v}$ and edges $\mathbf{e}$ of the $N$ equilateral triangles composing a regular hexagonal aperture $A$ (Fig. 2). The orientation of the edges is assumed clockwise for the point-down triangles and counterclockwise for the point-up triangles, that are colored in black and white in Fig. 2, respectively. Moreover, the notation $e_{i\rightarrow g}$ is used to indicate an edge connecting two adjacent vertices $v_i$ and $v_g$, oriented from $v_i$ towards $v_g$. As it can be inferred from Fig. 2, it turns out that

- a tile, whatever its orientation, namely vertical $\sigma^V$, horizontal-left $\sigma^L$, or horizontal-right $\sigma^R$, is obtained by the union of two triangles sharing one side (i.e., a diamond tile is the combination of a black and a white triangle);
- the triangles having one side on the boundary $\partial A$ of the aperture $A$ can generate only two different tile shapes. For example, the white triangles on the bottom of $A$ with the edges $e_{1\rightarrow M}$ and $e_{M\rightarrow M-1}$ on $\partial A$ (Fig. 2) can be included into a $\sigma^L$ or a $\sigma^R$ tile, but not into a $\sigma^V$ tile. Differently, all other triangles within $A$, having all sides shared with another triangle, can be potentially combined into one of the three admissible tiles in Fig. 1$(c)$.

As for the *HF* values of the external vertices, $\mathbf{v}^{(ext)} = \left\{v_m^{(ext)} \in \partial A : m = 1, ..., M\right\}$ ($M \triangleq 6 \times N_\ell$), they are only function of the size of the hexagonal aperture and they do not depend on the $t$-th ($t = 1, ..., T$) tiling configuration. Therefore, they are computed once. Towards this end, let us order the external vertices clockwise starting from the bottom-left vertex of the aperture $A$ (Fig. 2). The *HF* value of the first vertex is set to zero, namely $h_1 = h\left(v_1^{(ext)}\right) = 0$, and that of the others ($m = 2, ..., M$) is iteratively computed (Fig. 19) according to the following rules

$$\begin{array}{ll} h_{m+1} = h_m + 1 & if\ e_{m\rightarrow m+1} \\ h_{m+1} = h_m - 1 & if\ e_{m+1\rightarrow m}. \end{array} \quad (17)$$

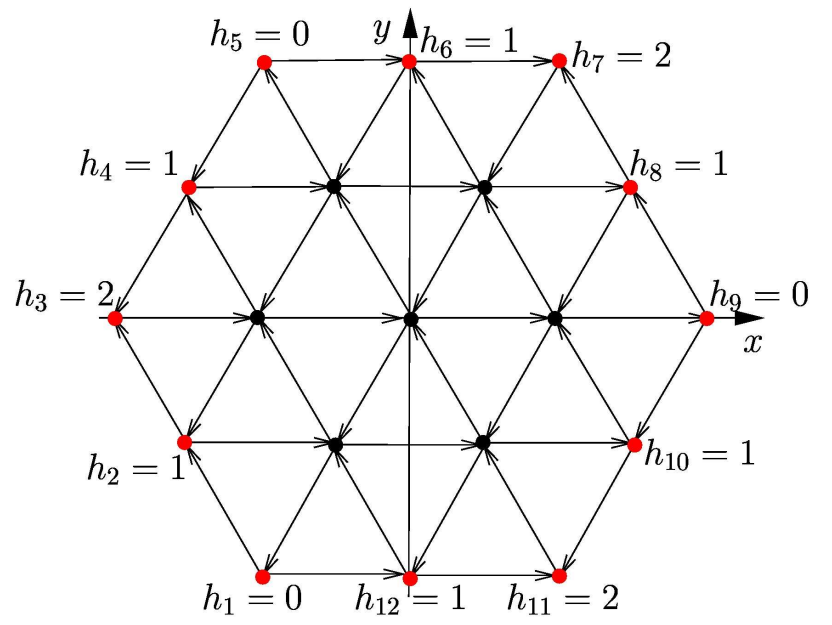

Figure 19. *Illustrative Geometry* ($N_\ell = 2$, $N = 24$) - *HF* values, $\{h_m$; $m = 1, ..., M\}$ of the external vertices, $\{v_m^{(ext)}$, $m = 1, ..., M\}$.

As for the internal vertices $\mathbf{v}^{(int)} = \left\{v_l^{(int)} : l = 1, ..., L\right\}$, they are indexed according to a raster order from bottom-left to top-right as indicated in Fig. 2. Unlike the external vertices, the *HF* values of the internal vertices are function of the tiling configuration (i.e., $h_l^{(t)} = h\left(v_l^{(int)}\right) = h\left(\mathbf{c}^{(t)}\right)$, $l = 1, ..., L$). Each value $h_l^{(t)}$ ($l = 1, ..., L$) is computed by first selecting a neighboring vertex $v_g \in \left\{\mathbf{v}^{(ext)}, \mathbf{v}^{(int)}\right\}$ with *HF* value already assigned[3] and then using the following four options

$$\begin{array}{lll} if\ e_{l\rightarrow g}\, and\, e_{l\rightarrow g} \in \partial\sigma_q & \Rightarrow & h_l^{(t)} = h_g^{(t)} + 1 \\ if\ e_{g\rightarrow l}\, and\, e_{g\rightarrow l} \in \partial\sigma_q & \Rightarrow & h_l^{(t)} = h_g^{(t)} - 1 \\ if\ e_{l\rightarrow g}\, and\, e_{l\rightarrow g} \notin \partial\sigma_q & \Rightarrow & h_l^{(t)} = h_g^{(t)} + 2 \\ if\ e_{g\rightarrow l}\, and\, e_{g\rightarrow l} \notin \partial\sigma_q & \Rightarrow & h_l^{(t)} = h_g^{(t)} - 2 \end{array} \quad (18)$$

where $\partial\sigma_q$ is the boundary of a diamond tile ($q = 1, ..., Q$) inside $A$. In (18), the condition $e_{l\rightarrow g} \in \partial\sigma_q$ means that the edge $e_{l\rightarrow g}$ belongs to the contour of a tile, while $e_{l\rightarrow g} \notin \partial\sigma_q$ indicates that the edge is covered by a tile.

## *Appendix IV* - Minimum Tiling

Unlike the domino-tiling for rectangular apertures and thanks to the regularity of the hexagonal aperture $A$, the *minimal tiling*, $\mathbf{c}^{(1)}$, can be simply yielded by filling the three parts of $A$ (i.e. the vertical, the horizontal-left, and the horizontal-right rhombus highlighted with different colors in Fig. 4) with vertical, $\sigma^V$, horizontal-left, $\sigma^L$, and horizontal-right, $\sigma^L$, tiles, respectively (Fig. 4).

By definition, the *minimal tiling word* is the first word, $\mathbf{w}^{(1)}$, and it has all entries equal to zero ($w_l^{(1)} = 0$, $l = 1, ..., L$), $\mathbf{w}^{(1)} = \mathbf{0}$ .

The values of the *HF* of the *minimum tiling*, $\{h_l^{(1)}$; $l = 1, ..., L\}$, are computed [Fig. 3$(a)$ - right plot, $\mathbf{h}^{(1)}$] according to *Appendix III*.

[3] The first internal vertices to be considered will be those connected to the external ones $\mathbf{v}^{(ext)}$ because the *HF* values of these latter can be computed once known the array size without assuming any clustering configuration.

### *Appendix V* - Thurston's Algorithm

The Thurston's Algorithm [36] for the computation of the *HF* values of the vertices on $\partial\tilde{A}$ is applied according to the following procedural steps:

- **Step 1.** *Vertex Selection* - Select the vertex on $\partial\tilde{A}$ with maximum *HF* value. If multiple vertices have the same maximum value, arbitrarily select one of them;
- **Step 2.** *Tile Placement* - Place one of the available diamond tiles ($\sigma^V$ or $\sigma^L$ or $\sigma^R$) so that the two vertices on $\partial\tilde{A}$ adjacent to the vertex selected in *Step 1* are also vertices of the diamond tile;
- **Step 3.** *Aperture Boundary and HF Update* - Complete the computation of *HF* on the vertices of the last placed diamond tile according to the rules defined in (18) and update the aperture boundary $\partial\tilde{A} \leftarrow \partial\left(\tilde{A} - \sigma^{V/L/R}\right)$ by subtracting the area of the newly placed tile;
- **Step 4.** *Termination Criterion* - Stop if the aperture is totally covered and the *HF* is computed for all the internal vertices of $A$. Otherwise, go to the *Step* 1.

## Acknowledgements

The authors thank Prof. A. D. Capobianco (University of Padova, Italy) for providing the data from CST Microwave Studio full-wave simulations. A. Massa wishes to thank E. Vico for her never-ending inspiration, support, guidance, and help.